\documentclass[aps,twocolumn,showpacs,letterpaper]{revtex4}
\usepackage{graphicx}
\usepackage{dcolumn}
\usepackage{amsmath}

\begin{document}

\title{Deformation effects and neutrinoless positron $\beta \beta $ decay of
$^{96}$Ru, $^{102}$Pd, $^{106}$Cd, $^{124}$Xe, $^{130}$Ba and $^{156}$Dy
isotopes within Majorona neutrino mass mechanism}
\author{P. K. Rath$^{1}$, R. Chandra$^{1,2}$, K. Chaturvedi$^{1,3}$,
P. K. Raina$^{2} $ and J. G. Hirsch$^{4}$}

\affiliation{
$^{1}$Department of Physics, University of Lucknow, Lucknow-226007, India\\
$^{2}$Department of Physics and Meteorology, IIT, Kharagpur-721302, India\\
$^{3}$Department of Physics, Bundelkhand University, Jhansi-284128, India\\
$^{4}$Instituto de Ciencias Nucleares, Universidad Nacional Aut\'{o}noma de
M\'{e}xico, A.P. 70-543, M\'{e}xico 04510 D.F., M\'{e}xico }
\date{\today}

\begin{abstract}
The $\left( \beta ^{+}\beta ^{+}\right) _{0\nu }$ and $\left( \varepsilon
\beta ^{+}\right) _{0\nu }$ modes of $^{96}$Ru, $^{102}$Pd, $^{106}$Cd, $%
^{124}$Xe, $^{130}$Ba and $^{156}$Dy isotopes are studied in the Projected
Hartree-Fock-Bogoliubov framework for the $0^{+}\rightarrow 0^{+}$
transition. The reliability of the intrinsic wave functions required to
study these decay modes has been established in our earlier works by obtaining an 
overall agreement between the theoretically calculated spectroscopic properties, 
namely yrast spectra,
reduced $B(E2$:$0^{+}\rightarrow 2^{+})$ transition probabilities,
quadrupole moments $Q(2^{+})$ and gyromagnetic factors $g(2^{+})$ and the
available experimental data in the parent and daugther even-even nuclei.
In the present work, the required nuclear
transition matrix elements are calculated in the Majorana neutrino mass 
mechanism using the
same set of intrinsic wave functions as used to study the two neutrino positron 
double-$\beta$ decay modes.
Limits on effective light neutrino mass $%
\left\langle m_{\nu }\right\rangle $ and effective heavy neutrino mass $%
\left\langle M_{N}\right\rangle $ are extracted from the observed limits on
half-lives $T_{1/2}^{0\nu }(0^{+}\rightarrow 0^{+})$ of $\left( \beta
^{+}\beta ^{+}\right) _{0\nu }$ and $\left( \varepsilon \beta ^{+}\right)
_{0\nu }$ modes. We also investigate the effect of quadrupolar correlations
vis-a-vis deformation on NTMEs required to study the $\left( \beta ^{+}\beta
^{+}\right) _{0\nu }$ and $\left( \varepsilon \beta ^{+}\right) _{0\nu }$
modes.
\end{abstract}


\maketitle

\section{INTRODUCTION}

The sixteen rare, experimentally distinguishable, modes of nuclear $\beta
\beta $ decay, namely the double-electron emission $\left( \beta ^{-}\beta
^{-}\right) $, double-positron emission $\left( \beta ^{+}\beta ^{+}\right) $%
, electron-positron conversion $\left( \varepsilon \beta ^{+}\right) $ and
double-electron capture $\left( \varepsilon \varepsilon \right) $ with the
emission of two neutrinos, no neutrinos, single Majoron and double Majorons,
are semileptonic weak transitions involving strangeness conserving charged
currents. The $\beta ^{+}\beta ^{+}$, $\varepsilon \beta ^{+}$ and $%
\varepsilon \varepsilon $ modes are energetically competing and we shall
refer to them as $e^{+}\beta \beta $ decay. 
The experimental as well as theoretical study of nuclear $\beta^- \beta^- $ 
mode has been excellently reviewed over the 
past decades, which can be found in the recent review \cite{avig08} and 
references there in. Also, the experimental and
theoretical studies devoted to the $e^{+}\beta \beta $ decay have been
reviewed over the past years \cite
{rose65,verg83,verg86,doi92,doi93,bara95,suho98,kirp00,klap01,bara04}. Owing
to the confirmation of flavour oscillation of neutrinos at atmospheric,
solar, reactor and accelerator neutrino sources, it has been established that
neutrinos have mass. However, it is generally agreed that the observation of 
$\left( \beta \beta \right) _{0\nu }$ decay can clarify a number of issues
regarding the nature of neutrinos, namely the origin of neutrino mass (Dirac
vs. Majorana), the absolute scale on neutrino mass, the type of hierarchy and CP
violation in the leptonic sector, etc. Further, the possible mechanisms for the
occurrence of the lepton number violating $\left( \beta \beta \right) _{0\nu
}$ decay are the exchange of light as well as heavy neutrinos and the right
handed currents in the LRSM, the exchange of sleptons, neutralinos, squarks
and gluinos in the $R_{p}$-violating MSSM, the exchange of leptoquarks,
existence of heavy sterile neutrinos, compositeness and extradimensional
scenarios. In nine Majoron models, namely $IB$, $IC$, $IIB$, $IIC$, $IIF$, 
$ID$, $IE$, $IID$ and $IIE$ \cite{bame95},
the single Majoron accompanied neutrinoless
double beta $\left( \beta \beta \phi \right) _{0\nu }$ decay and double
Majoron accompanied neutrinoless double beta $\left( \beta \beta \phi \phi
\right) _{0\nu }$ decay occur in 
the former five and the latter four, respectively. 
The study of $\left( \beta \beta \right) _{0\nu }$ decay can
provide stringent limits on the associated gauge theoretical parameters and
its observation can only ascertain the role of various 
possible mechanisms in different gauge theoretical models.

In principle, the $\beta ^{-}\beta ^{-}$ decay and $e^{+}\beta \beta $ decay
can provide us with the same but complementary information. The observation
of $\left( e^{+}\beta \beta \right) _{2\nu }$ decay modes will be interesting from
the nuclear structure point of view, as it 
is
 a challenging task to
calculate the nuclear transition matrix elements (NTMEs) of these modes along 
with $\left( \beta ^{-}\beta
^{-}\right) _{2\nu }$ mode in the same theoretical framework. Further, the
observation of $\left( e^{+}\beta \beta \right) _{0\nu }$ decay modes will
be helpful in deciding issues like dominance of mass mechanism or right
handed currents \cite{hirs94}. In an attempt to study the role of $m_{\nu }$%
, $\lambda $ and $\eta $ mechanisms, Klapdor-Kleingrothaus \textit{et al.}
have analyzed the 71.7 kg.y data collected from 1990-2003 on enriched $^{76}$%
Ge \cite{klap04} and have shown that there is an apparent degeneracy in the
parameters \cite{klap06}. 
It has been also concluded that the analysis of a
high sensitive $\left( \beta ^{-}\beta ^{-}\right) _{0\nu }$ experiment e.g. 
$^{76}$Ge and a suitable high sensitive mixed mode decay e.g. $^{124}$Xe is
more advantageous \cite{hirs94}.

In spite of the fact that the kinetic energy release in the $\left( \varepsilon
\varepsilon \right) _{0\nu }$ mode is the largest, the experimental and
theoretical study of this mode has not been attempted so far. The
conservation of energy-momentum requires the emission of an additional
particle in the $\left( \varepsilon \varepsilon \right) _{0\nu }$ mode.
Further, the emission of one real photon is forbidden for the $%
0^{+}\rightarrow 0^{+}$ transition if atomic electrons are absorbed from the 
$K$-shell. Therefore, one has to consider various processes such as internal
pair production, internal conversion, emission of two photons, $L$-capture
etc. \cite{doi93}. The decay rates of the above mentioned processes have to
be calculated at least by the third order perturbation theory. Resultingly,
there is a suppression factor of the order of 10$^{-4}$ 
in comparison to the $\left( \varepsilon \beta ^{+}\right) _{0\nu }$ mode. Hence, the experimental
as well as theoretical study of $\left( e^{+}\beta \beta \right) _{0\nu }$
decay has been restricted to $\left( \beta ^{+}\beta ^{+}\right) _{0\nu }$
and $\left( \varepsilon \beta ^{+}\right) _{0\nu }$ modes only. Arguably,
Sujkowski and Wycech \cite{sujo04} have shown that there will be resonant
enhancement of the $\left( \varepsilon \varepsilon \right) _{0\nu }$ mode if the
nuclear levels in parent and daughter nuclei are almost degenerate i.e. $%
Q-(E_{2P}-E_{2S})$ $\sim 1$ $keV$, where the energy difference is for
atomic levels. Interestingly, Barabash $et$ $al.$ have reported that there
might be a degeneracy between the $^{112}$Sn
ground state
 and an excited 0$^{+}$ state at
1870.9 keV in $^{112}$Cd fulfilling the resonance enhancement condition for
the $\left( \varepsilon \varepsilon \right) _{0\nu }$ mode \cite{bara08}.
It is expected that the study of this $\left( \varepsilon \varepsilon \right)
_{0\nu }$ mode may be interesting in the near future.

The complex structure of nuclei in general, and of mass region $96<A<156$ in
particular, is due to the subtle interplay of pairing and multipolar
correlations present in the effective two-body interaction. The mass regions 
$A\sim 100$ and $150$ offer nice examples of shape transitions at $N=60$ 
and $90$, respectively. 
The nuclei are soft vibrators for neutron number $N<60$ and $N<90$ and 
quasi-rotors for $N>60$ and $N>90$. Nuclei with neutron numbers $N=60$ 
and $90$ are transitional nuclei.
The yrast spectra of Te and Xe isotopes, on the other hand, follow an
approximate inverse parabolic type of systematics with minimum energy of
$2^+$ states occurring for $^{120}$Te and $^{120}$Xe isotopes, respectively. 
In this
mass region $96<A<156,$ the deformation parameters $\beta _{2}$ are in the
range $(0.1409\pm 0.0046)-(0.3378\pm 0.0018)$ corresponding to 
$^{132}$Xe and $^{156}$Gd isotopes, respectively and hence, 
it is clear that deformation
plays a crucial role in reproducing the properties of these nuclei. In
nuclear $\beta \beta $ decay, the role of deformation degrees of freedom in
addition to pairing correlation has been already stressed \cite
{grif92,suho94}. Recently, the effects of pairing and quadrupolar
correlations on the NTMEs of $\left( \beta ^{-}\beta ^{-}\right) _{0\nu }$
mode has been studied in the ISM \cite{caur08,mene08}. In the PHFB model,
the role of deformation effects due to quadrupolar \cite
{chan05,rain06,sing07,chat08} and multipolar correlations \cite{chan09} has
been also studied.

The shell model is the best choice for calculating the NTMEs as it attempts
to solve the nuclear many-body problem as exactly as possible. However, the
first explanation about the observed suppression of $M_{2\nu }$ was provided
in the QRPA model by Vogel and Zirnbauer \cite{voge87} and Civitarese $et$ $%
al.$ \cite{civi87}. Further, the QRPA and its extensions have emerged as the
most successful models in correlating single-$\beta $ GT strengths and
half-lives of ($\beta ^{-}\beta ^{-}$)$_{2\nu }$ mode. In spite of the
spectacular success of the QRPA in the study of $\beta \beta $ decay, the
necessity to include the deformation degrees of freedom in its formalism led
to the development of the deformed QRPA model for studying $\beta \beta $
decay of spherical as well as deformed nuclei. The effect of deformation on
the $\left( \beta ^{-}\beta ^{-}\right) _{2\nu }$ mode for the ground state
transition $^{76}$Ge $\rightarrow $ $^{76}$Se was studied in the framework
of deformed QRPA with separable GT residual interaction \cite{pace04} 
and, very recently, employing realistic forces \cite{you09}. 
A deformed QRPA formalism to describe simultaneously the energy distributions
of the single-$\beta $ GT strength and the $\left( \beta ^{-}\beta
^{-}\right) _{2\nu }$ mode matrix elements for $^{48}$Ca, $^{76}$Ge, $^{82}$%
Se, $^{96}$Zr, $^{100}$Mo, $^{116}$Cd, $^{128,130}$Te, $^{136}$Xe and $%
^{150} $Nd isotopes using deformed Woods-Saxon potential and deformed
Skyrme Hartree-Fock mean field was developed \cite{alva04}. Rodin and
Faessler \cite{rodi08} have studied the $\beta^{-} \beta^{-} $ decay 
of $^{76}$Ge, $%
^{100}$Mo and $^{130}$Te isotopes and it has been reported that the effect
of continuum on the NTMEs of $\left( \beta ^{-}\beta ^{-}\right) _{2\nu }$
mode is negligible whereas the NTMEs of $\left( \beta ^{-}\beta ^{-}\right)
_{0\nu }$ mode are regularly suppressed.

In the PHFB model, the interplay of pairing and deformation degrees of
freedom are treated simultaneously and on equal footing. However, the
structure of the intermediate odd $Z$-odd $N$ nuclei, which provide
information on the single-$\beta $ decay rates and the distribution of GT
strengths, can not be studied in the present version of the PHFB model. In
spite of this limitation, the PHFB model, in conjunction with pairing plus
quadrupole-quadrupole (\textit{PQQ}) \cite{bara68} interaction has been
successfully applied to study the $0^{+}\rightarrow 0^{+}$ transition of $%
\left( \beta ^{-}\beta ^{-}\right) _{2\nu }$ mode, where it was possible to
describe the lowest excited states of the parent and daughter nuclei along
with their electromagnetic transition strengths, as well as to reproduce
their measured $\beta ^{-}\beta ^{-}$ decay rates \cite{chan05,sing07}. The
main purpose of using the \textit{PQQ} interaction is to study the interplay
between sphericity and deformation. In this way, the PHFB formalism,
employed in conjunction with the \textit{PQQ} interaction, is a convenient
choice to examine the explicit role of deformation on the NTMEs. The
existence of an inverse correlation between the quadrupole deformation and
the size of NTME $M_{2\nu }$ has been also confirmed \cite
{chan05,rain06,sing07}. In addition, it has been observed that the NTMEs for 
$\beta ^{-}\beta ^{-}$ decay are usually large in the absence of quadrupolar
correlations. With the inclusion of the quadrupolar correlations, the NTMEs
are almost constant for small admixture of the $QQ$ interaction and
suppressed substantially in realistic situation. It was also shown that the
NTMEs of $\beta ^{-}\beta ^{-}$ decay have a well defined maximum when the
deformation of parent and daughter nuclei are similar and they are
suppressed for a difference in deformations
in agreement with previous QRPA calculations \cite{pace04}. 
The deformation effects are
also of equal importance in the case of $\left( \beta ^{-}\beta ^{-}\right)
_{2\nu }$ and $\left( \beta ^{-}\beta ^{-}\right) _{0\nu }$ modes \cite
{chat08,chan09}.

Moreover, the PHFB model along with the \textit{PQQ} interaction in
conjunction with the summation method has been successfully applied to study
the $\left( e^{+}\beta \beta \right) _{2\nu }$ decay of $^{96}$Ru, $^{102}$%
Pd, $^{106,108}$Cd, $^{124,126}$Xe, $^{130,132}$Ba \cite{rain06,sing07} and $%
^{156}$Dy \cite{rath09} isotopes for the $0^{+}\to 0^{+}$ transition, not in
isolation but together with other observed nuclear spectroscopic properties,
namely yrast spectra, reduced $B(E2$:$0^{+}\rightarrow 2^{+})$ transition
probabilities, quadrupole moments $Q(2^{+})$ and gyromagnetic factors $%
g(2^{+})$. This success of the PHFB model has prompted us to apply the same
to study the $0^{+}\rightarrow 0^{+}$ transition of 
$\left( \beta ^{+}\beta ^{+}\right) _{0\nu }$ and $\left(
\varepsilon \beta ^{+}\right) _{0\nu }$ modes for the above mentioned nuclei. 
It has been observed that in
general, there exists an anticorrelation between
the magnitude of 
 the quadrupolar deformation
and the NTMEs $M_{2\nu }$ of $\left( e^{+}\beta \beta \right) _{2\nu }$
decay. In the case of $\left( e^{+}\beta \beta \right) _{2\nu }$ decay, we
observed that the deformation plays an important role in the suppression of $%
M_{2\nu }$ by a factor of 2--13.6 approximately \cite{rain06,sing07,rath09}%
. Therefore, we aim to study the variation of NTMEs of $\left( \beta
^{+}\beta ^{+}\right) _{0\nu }$ and $\left( \varepsilon \beta ^{+}\right)
_{0\nu }$ modes vis-a-vis the change in deformation by
changing
the
strength of the \textit{QQ} interaction.

The present paper is organized as follows. The theoretical formalism for
calculating the half-lives of $\left( \beta ^{+}\beta ^{+}\right) _{0\nu }$
and $\left( \varepsilon \beta ^{+}\right) _{0\nu }$ modes has been given by
Doi \textit{et al. }\cite{doi93}. Hence, we briefly outline steps of the
detailed derivations in Sec. II. In Sec. III, we present the results and
discuss them vis-a-vis the existing calculations done in other nuclear
models. In the study of $\left( \beta \beta \right) _{0\nu }$ decay, the
practice is to either extract limits on various gauge theoretical parameters
from the observed limits on half-lives of the $\left( \beta \beta \right)
_{0\nu }$ decay or predict half-lives assuming certain value for the
neutrino mass. Presently, the available experimental limits on half-lives of 
$\left( \beta ^{+}\beta ^{+}\right) _{0\nu }$ and $\left( \varepsilon \beta
^{+}\right) _{0\nu }$ modes are not large enough to provide stringent limits
on the effective gauge theoretical parameters $\left\langle m_{\nu
}\right\rangle $ and $\left\langle M_{N}\right\rangle $. Therefore, we also
predict half-lives $T_{1/2}^{0\nu }(0^{+}\rightarrow 0^{+})$ of $\left(
\beta ^{+}\beta ^{+}\right) _{0\nu }$ and $\left( \varepsilon \beta
^{+}\right) _{0\nu }$ modes for $^{96}$Ru, $^{102}$Pd, $^{106}$Cd, $^{124}$%
Xe, $^{130}$Ba and $^{156}$Dy isotopes, which will be helpful in the future
experimental studies of $\left( e^{+}\beta \beta \right) _{0\nu }$ decay. 
In addition, we study the deformation effect on NTMEs of $\left(
\beta ^{+}\beta ^{+}\right) _{0\nu }$ and $\left( \varepsilon \beta
^{+}\right) _{0\nu }$ modes and show that the NTMEs have well defined
maximum for similar deformations of parent and daughter nuclei and they are
suppressed for a difference in deformations. Finally, the conclusions are
given in Sec. IV.

\section{THEORETICAL FORMALISM}

In the Majorana neutrino mass mechanism, 
the effective charged current weak interaction
Hamiltonian density $H_{W}$ for $\beta ^{+}$ decay due to $W$-boson exchange
including hadronic currents can be written as

\begin{equation}
H_{W}=\frac{G}{\sqrt{2}}j_{L\mu }J_{L}^{\mu \dagger }+h.c..
\end{equation}

\noindent The left handed \textit{V }$-$\textit{\ A }leptonic and hadronic
currents for $\beta ^{+}$ decay are given by

\begin{eqnarray}
j_{L}^{\mu } &=&\overline{\nu _{eL}}\gamma ^{\mu }\left( 1-\gamma
_{5}\right) e, \\
J_{L}^{\mu \dagger } &=&g_{v}\overline{d}\gamma ^{\mu }\left( 1-\gamma
_{5}\right) u,
\end{eqnarray}

\noindent where $g_{v}=\cos \theta _{c}$ and $\theta _{c}$ is the
Cabibbo-Kobayashi-Maskawa (CKM) mixing angle for the left and right handed $%
d $ and $s$ quarks. Further,

\begin{equation}
\nu _{eL}=\sum_{i}\text{ }U_{ei}N_{iL}.  \label{mix}
\end{equation}
The Majorana neutrino field $N_{i}$ has mass $m_{i}$ and the mixing matrices 
$U$ of left handed neutrinos are normalized i.e. $\sum\limits_{i}\left|
U_{ei}\right| ^{2}=1.$

Usually, the decay rates for the $0^{+}\to 0^{+}$\ transition of $\left(
\beta ^{+}\beta ^{+}\right) _{0\nu }$ and $\left( \varepsilon \beta
^{+}\right) _{0\nu }$ modes are derived by making the following assumptions:

\noindent (i) The light and heavy neutrino species of mass $m_{i}<10$ eV and 
$m_{i}>1$ GeV, respectively are only considered.

\noindent (ii) The nonrelativistic impulse approximation is assumed for the
hadronic currents.

\noindent (iii) The recoil current is neglected. However, it has been shown
by \v{S}imkovic $et$ $al$. \cite{simk99} and Vergados \cite{verg02} that the
consideration of pseudoscalar and weak magnetism terms of recoil current
reduce the NTMEs up to 30\%, which needs to be further investigated.

\noindent (iv) The \textit{s}$_{1/2}$ waves describe the final leptonic
states.

\noindent (v) 
The calculation of phase space factors is made easier by considering
no finite de Broglie wave length correction.

\noindent (vi) The CP conservation is assumed. Consequently, the effective
light neutrino mass $\left\langle m_{\nu }\right\rangle $ and effective heavy
neutrino mass $\left\langle M_{N}\right\rangle $ are real.

With these approximations, the inverse half-lives $T_{1/2}^{0\nu }$
for the $0^{+}\rightarrow 0^{+}$ transition of 
$\left( \beta ^{+}\beta ^{+}\right) _{0\nu }$ and $\left( \varepsilon
\beta ^{+}\right) _{0\nu }$ modes in 2n mechanism are given by \cite{doi93}
\begin{widetext}
\begin{eqnarray}
\left[ T_{1/2}^{0\nu }\left( \beta \right) \right] ^{-1} &=&\left( \frac{%
\left\langle m_{\nu }\right\rangle }{m_{e}}\right) ^{2}G_{01}\left( \beta
\right) \left( M_{GT}-M_{F}\right) ^{2}+\left( \frac{m_{p}}{\left\langle
M_{N}\right\rangle }\right) ^{2}G_{01}\left( \beta \right) \left(
M_{GTh}-M_{Fh}\right) ^{2}  \nonumber \\
&&+\left( \frac{\left\langle m_{\nu }\right\rangle }{m_{e}}\right) \left( 
\frac{m_{p}}{\left\langle M_{N}\right\rangle }\right) G_{01}\left( \beta
\right) \left( M_{GT}-M_{F}\right) \left( M_{GTh}-M_{Fh}\right) ,
\end{eqnarray}
\end{widetext}
\noindent where $\beta $ denotes the $\left( \beta ^{+}\beta ^{+}\right)
_{0\nu }/\left( \varepsilon \beta ^{+}\right) _{0\nu }$ mode and 
\begin{eqnarray}
\left\langle m_{\nu }\right\rangle &=&\sum\nolimits_{i}^{\prime
}U_{ei}^{2}m_{i},\qquad \qquad m_{i}<10\text{ }eV, \\
\left\langle M_{N}\right\rangle ^{-1} &=&\sum\nolimits_{i}^{\prime \prime
}U_{ei}^{2}m_{i}^{-1},\qquad \qquad m_{i}>1\text{ }GeV.
\end{eqnarray}

\noindent In the closure approximation, NTMEs $M_{F}$, $M_{GT}$, $M_{Fh}$
and $M_{GTh}$ are written as 
\begin{widetext}
\begin{eqnarray}
M_{F} &=&\left( \frac{g_{V}}{g_{A}}\right) ^{2}\sum_{n,m}\left\langle
0_{F}^{+}\left\| H(r)\tau _{n}^{-}\tau _{m}^{-}\right\|
0_{I}^{+}\right\rangle , \\
M_{GT} &=&\sum_{n,m}\left\langle 0_{F}^{+}\left\| \mathbf{\sigma }_{n}\cdot 
\mathbf{\sigma }_{m}H(r)\tau _{n}^{-}\tau _{m}^{-}\right\|
0_{I}^{+}\right\rangle , \\
M_{Fh} &=&4\pi \left( M_{p}m_{e}\right) ^{-1}\left( \frac{g_{V}}{g_{A}}%
\right) ^{2}\sum_{n,m}\left\langle 0_{F}^{+}\left\| \delta \left( \mathbf{r}%
\right) \tau _{n}^{-}\tau _{m}^{-}\right\| 0_{I}^{+}\right\rangle , \\
M_{GTh} &=&4\pi \left( M_{p}m_{e}\right) ^{-1}\sum_{n,m}\left\langle
0_{F}^{+}\left\| \mathbf{\sigma }_{n}\cdot \mathbf{\sigma }_{m}\delta \left( 
\mathbf{r}\right) \tau _{n}^{-}\tau _{m}^{-}\right\| 0_{I}^{+}\right\rangle .
\end{eqnarray}
\end{widetext}
The neutrino potential $H(r)$ arising due to the exchange of light
neutrino is defined as 
\begin{equation}
H\left( r\right) =\frac{4\pi R}{\left( 2\pi \right) ^{3}}\int d^{3}q\frac{%
\exp \left( i\mathbf{q}\cdot \mathbf{r}\right) }{\omega \left( \omega +%
\overline{A}\right) },
\end{equation}
with 
\begin{equation}
\overline{A}=\left\langle E_{N}\right\rangle -\frac{1}{2}\left(
E_{I}+E_{F}\right) .
\end{equation}
\qquad In addition, the inclusion of effects due to finite size of nucleons
(FNS) and short range correlations (SRC) is required. The FNS is usually
taken into account by a dipole type of form factor making the replacement

\begin{equation}
g_{V}\rightarrow g_{V}\left( \frac{\Lambda ^{2}}{\Lambda ^{2}+k^{2}}\right)
^{2}\qquad \text{and}\qquad g_{A}\rightarrow g_{A}\left( \frac{\Lambda ^{2}}{%
\Lambda ^{2}+k^{2}}\right) ^{2}
\end{equation}
with $\Lambda =850$ MeV. In the PHFB model, the configuration mixing takes care
of the long range correlations. The effect of short range correlations
(SRC), which arise mainly from the repulsive nucleon-nucleon potential due
to the exchange of $\rho $ and $\omega $ mesons, is usually absent. To study 
the $\left( \beta ^{-}\beta ^{-}\right) _{0\nu }$ mode, the SRC
has been incorporated by Hirsch \textit{et al. } through the exchange of $%
\omega $-meson \cite{jghi95}, Kortelainen \textit{et al. }%
\cite{kort07} as well as \v{S}imkovic \textit{et al.} \cite{simk08} by using the
unitary correlation operator method (UCOM) and 
\v{S}imkovic \textit{et al.} \cite{simk09} by self-consistent CCM. 
This SRC effect can also be incorporated
through phenomenological Jastrow type of correlation using Miller and
Spencer parametrization by the prescription

\begin{equation}
\left\langle j_{1}^{\pi }j_{2}^{\pi }J\left| O\right| j_{1}^{\nu }j_{2}^{\nu
}J^{^{\prime }}\right\rangle \rightarrow \left\langle j_{1}^{\pi }j_{2}^{\pi
}J\left| fOf\right| j_{1}^{\nu }j_{2}^{\nu }J^{^{\prime }}\right\rangle ,
\end{equation}
where

\begin{equation}
f(r)=1-e^{-ar^{2}}(1-br^{2})
\end{equation}
with $a$ = 1.1 fm$^{-2}$ and $b$ = 0.68 fm$^{-2}$ \cite{mill76}. It has been
shown by Wu and co-workers \cite{wu85} that for the $\left( \beta ^{-}\beta
^{-}\right) _{0\nu }$ mode of $^{48}$Ca, the phenomenologically determined $%
f(r)$ has strong two nucleon correlations in comparison to the effective
transition operator $\widehat{f}O\widehat{f}$ derived using Reid and Paris
potentials.

In the PHFB model, the calculation of the NTMEs $M_{\alpha } \,
(\alpha=F, \; GT, \; Fh \; \rm{and} \; GTh)$ of the $%
\left( \beta ^{+}\beta ^{+}\right) _{0\nu }$ and $\left( \varepsilon \beta
^{+}\right) _{0\nu }$ modes is carried out as follows. The two basic
ingredients of the PHFB model are the existence of an independent
quasiparticle mean field solution and the projection technique. To start
with, amplitudes $(u_{im},v_{im})$ and expansion coefficients $C_{ij,m}$
required to specify the axially symmetric HFB intrinsic state ${|\Phi
_{0}\rangle }$ with $K=0$ are obtained by carrying out the HFB calculation
through the minimization of the expectation value of the effective
Hamiltonian. Subsequently, states with good angular momentum $\mathbf{J}$
are obtained from ${|\Phi _{0}\rangle }$ using the standard projection
technique \cite{onis66} given by 
\begin{equation}
{|\Psi _{00}^{J}\rangle }=\frac{(2J+1)}{{8\pi ^{2}}}\int D_{00}^{J}(\Omega
)R(\Omega )|\Phi _{0}\rangle d\Omega ,
\end{equation}
where $\ R(\Omega )$\ and $\ D_{00}^{J}(\Omega )$\ are the rotation operator
and the rotation matrix, respectively. Further,
\begin{equation}
{|\Phi _{0}\rangle }=\prod\limits_{im}(u_{im}+v_{im}b_{im}^{\dagger }b_{i%
\bar{m}}^{\dagger })|0\rangle
\end{equation}
with the creation operators $\ b_{im}^{\dagger }$\ and $\ b_{i\bar{m}%
}^{\dagger }$\ defined as 
\begin{equation}
b{_{im}^{\dagger }}=\sum\limits_{\alpha }C_{i\alpha ,m}a_{\alpha m}^{\dagger
}\;\, \hbox{and}\mathrm{\;\,}b_{i\bar{m}}^{\dagger }=\sum\limits_{\alpha
}(-1)^{l+j-m}C_{i\alpha ,m}a_{\alpha ,-m}^{\dagger }.
\end{equation}
Finally, the NTMEs $M_{\alpha }$ of the $\left( \beta ^{+}\beta ^{+}\right)
_{0\nu }$ and $\left( \varepsilon \beta ^{+}\right) _{0\nu }$ modes are
given by

\begin{eqnarray}
M_{\alpha } &=&\langle \Psi {_{00}^{J_{f}=0}}||O_{\alpha }\tau ^{-}\tau
^{-}||\Psi {_{00}^{J_{i}=0}}\rangle  \nonumber \\
&=&[n_{Z,N}^{J_{i}=0}n_{Z-2,N+2}^{J_{f}=0}]^{-1/2} \nonumber \\
&&\times \int\limits_{0}^{\pi
}n_{(Z,N),(Z-2,N+2)}(\theta )\sum\limits_{\alpha \beta \gamma \delta
}\left\langle \alpha \beta \left| O_{\alpha }\tau ^{-}\tau ^{-}\right|
\gamma \delta \right\rangle  \nonumber \\
&&\times \sum_{\varepsilon \eta }\frac{(f_{Z-2,N+2}^{(\nu )*})_{\varepsilon
\beta }}{\left[ 1+F_{Z,N}^{(\nu )}(\theta )f_{Z-2,N+2}^{(\nu )*}\right]
_{\varepsilon \alpha }} \nonumber \\
&&\times \frac{(F_{Z,N}^{(\pi )*})_{\eta \delta }}{\left[
1+F_{Z,N}^{(\pi )}(\theta )f_{Z-2,N+2}^{(\pi )*}\right] _{\gamma \eta }}\sin
\theta d\theta ,  \label{eqf}
\end{eqnarray}
where

\begin{eqnarray}
n^{J}&=&\int\limits_{0}^{\pi }\{\det [1+F^{(\pi )}(\theta )f^{(\pi )\dagger
}]\}^{1/2} \nonumber \\
&&\times \{\det [1+F^{(\nu )}(\theta )f^{(\nu )\dagger
}]\}^{1/2}d_{00}^{J}(\theta )\sin (\theta )d\theta \nonumber \\
&&
\end{eqnarray}
and

\begin{eqnarray}
n_{(Z,N),(Z-2,N+2)}(\theta )&=&\{\det [1+F_{Z,N}^{(\pi )}(\theta
)f_{Z-2,N+2}^{(\pi )\dagger }]\}^{1/2} \nonumber \\
&&\times \{\det [1+F_{Z,N}^{(\nu
)}(\theta )f_{Z-2,N+2}^{(\nu )\dagger }]\}^{1/2}. \nonumber \\
&&
\end{eqnarray}
The $\pi (\nu )$ represents the proton (neutron) of nuclei involved in the $%
\left( \beta ^{+}\beta ^{+}\right) _{0\nu }/\left( \varepsilon \beta
^{+}\right) _{0\nu }$ mode. The matrices $f_{Z,N}$\ \ and $F_{Z,N}(\theta )\ 
$are given by

\begin{equation}
\left[ f_{Z,N}\right] _{\alpha \beta }=\sum_{i}C_{ij_{\alpha },m_{\alpha
}}C_{ij_{\beta },m_{\beta }}\left( v_{im_{\alpha }}/u_{im_{\alpha }}\right)
\delta _{m_{\alpha },-m_{\beta }}  \label{eq1}
\end{equation}
and 
\begin{equation}
\left[ F_{Z,N}(\theta )\right] _{\alpha \beta }=\sum_{m_{\alpha }^{^{\prime
}}m_{\beta }^{^{\prime }}}d_{m_{\alpha },m_{\alpha }^{^{\prime
}}}^{j_{\alpha }}(\theta )d_{m_{\beta },m_{\beta }^{^{\prime }}}^{j_{\beta
}}(\theta )f_{j_{\alpha }m_{\alpha }^{^{\prime }},j_{\beta }m_{\beta
}^{^{\prime }}}.  \label{eq2}
\end{equation}
To calculate NTMEs $M_{\alpha }$ of the $\left( \beta ^{+}\beta ^{+}\right)
_{0\nu }$ and $\left( \varepsilon \beta ^{+}\right) _{0\nu }$ modes, the
matrices $\left[ f_{Z,N}\right] _{\alpha \beta }$ and $\left[ F_{Z,N}(\theta
)\right] _{\alpha \beta }$ are evaluated using expressions given by Eqs. (%
\ref{eq1}) and (\ref{eq2}), respectively. The required NTMEs $M_{\alpha }$
are obtained using Eq. (\ref{eqf}) with 20 gaussian quadrature points in the
range ($0$, $\pi $).

\section{RESULTS\ AND\ DISCUSSIONS}

The model space, single particle energies (SPE's) and parameters of the effective two-body
interaction are the same as our earlier calculations on $\left( e^{+}\beta
\beta \right) _{2\nu }$ decay of $^{96}$Ru, $^{102}$Pd, $^{106,108}$Cd \cite
{rain06}, $^{124,126}$Xe, $^{130,132}$Ba \cite{sing07} and $^{156}$Dy \cite
{rath09} isotopes for the $0^{+}\to 0^{+}$ transition. We briefly present a
discussion about them for the sake of completeness as well as present convenience.
The doubly even $^{76}$Sr ($N=Z=38$) and $^{100}$Sn ($N=Z=50$) nuclei were treated 
as inert cores for the nuclei in the mass region $A=96-108$ and $A=124-156$, respectively.
The change of model space was forced upon
because the number of neutrons increase to about 40 for nuclei occurring in
the mass region $A=130$ and with the increase in neutron number, the yrast
energy spectra was compressed due to increase in the attractive part of
effective two-body interaction. In Table~\ref{tab1}, we have given the single
particle orbits, which span the valence space and corresponding SPEs. In the
model space with $^{76}$Sr core, the 1\textit{p}$_{1/2}$ orbit was
included to examine the role of the $Z=40$ proton core vis-a-vis the onset
of deformation in the highly neutron rich isotopes. For $^{156}$Dy and $%
^{156}$Gd isotopes, the SPE's used for $0h_{11/2}$, $1f_{7/2}$ and $0h_{9/2}$
orbits were $4.6$ MeV, $11.0$ MeV and $11.6$ MeV, respectively.
\begin{table}[h]
\caption{Single particle orbits of the model space and SPEs for protons and
neutrons.}
\label{tab1}
\begin{tabular}{llcll}
\hline\hline
\multicolumn{2}{c}{$A=96-108$} & & \multicolumn{2}{c}{$A=124-156$} \\ 
Orbits\thinspace \thinspace \thinspace \thinspace \thinspace \thinspace \thinspace \thinspace \thinspace \thinspace \thinspace \thinspace \thinspace \thinspace \thinspace & $%
\varepsilon $ (MeV) & \multicolumn{1}{l}{}~~~~~ & Orbits\thinspace \thinspace \thinspace \thinspace \thinspace \thinspace \thinspace \thinspace \thinspace \thinspace \thinspace \thinspace \thinspace \thinspace
\thinspace \thinspace \thinspace \thinspace & $\varepsilon $ (MeV) \\ \hline
1\textit{p}$_{1/2}$ & $-0.8$ & \multicolumn{1}{l}{} & 2\textit{s}$_{1/2}$ & $%
1.4$ \\ 
2\textit{s}$_{1/2}$ & $\,\,\,\,6.4$ & \multicolumn{1}{l}{} & 1\textit{d}$%
_{3/2}$ & $2.0$ \\ 
1\textit{d}$_{3/2}$ & $\,\,\,\,7.9$ & \multicolumn{1}{l}{} & 1\textit{d}$%
_{5/2}$ & $0.0$ \\ 
1\textit{d}$_{5/2}$ & $\,\,\,\,5.4$ & \multicolumn{1}{l}{} & 1\textit{f}$%
_{7/2}$ & $12.0$ \\ 
0\textit{g}$_{7/2}$ & $\,\,\,\,8.4$ & \multicolumn{1}{l}{} & 0\textit{g}$%
_{7/2}$ & $4.0$ \\ 
0\textit{g}$_{9/2}$ & $\,\,\,\,0.0$ & \multicolumn{1}{l}{} & 0\textit{h}$%
_{9/2}$ & $12.5$ \\ 
0\textit{h}$_{11/2}$ & $\,\,\,\,8.6$ & \multicolumn{1}{l}{} & 0\textit{h}$%
_{11/2}$ & $6.5$ \\ \hline\hline
\end{tabular}
\end{table}

The HFB wave functions were generated by using an effective Hamiltonian with 
\textit{PQQ} type of effective two-body interaction \cite{bara68} given by 
\begin{equation}
H=H_{sp}+V(P)+\zeta _{qq}V(QQ),
\end{equation}
where $H_{sp}$, $V(P)$ and $V(QQ)$ represent the single particle
Hamiltonian, the pairing and quadrupole-quadrupole part of the effective
two-body interaction, respectively. The arbitrary parameter $\zeta _{qq}$ was
introduced to study the role of deformation by varying the strength of 
\textit{QQ} interaction and the final results were obtained by using $\zeta
_{qq}=1$. Following Heestand \textit{et al.} \cite{hees69}, who have used $G_{p}=30/A$ MeV
and $G_{n}=20/A$ MeV to explain the experimental $g(2^{+})$ data of some
even-even Ge, Se, Mo, Ru, Pd, Cd and Te isotopes in Greiner's collective
model \cite{grei66}, we used the same strengths for $A=96-108$ nuclei. In
the case of $A=124-132$ isotopes, the strengths of the pairing interaction
were fixed as $G_{p}=G_{n}=35/A$ MeV. However, we used $G_{p}=G_{n}=30/A$
MeV for $^{156}$Dy and $^{156}$Gd isotopes.

The parameters of the \textit{QQ} interaction were fixed as follows. The
strengths of the like particle components $\chi _{pp}$ and $\chi _{nn}$ were
taken as $0.0105$ MeV \textit{b}$^{-4}$, where \textit{b} is oscillator
parameter. The strength of proton-neutron (\textit{pn}) component $\chi
_{pn} $ was varied so as to obtain the spectra of considered nuclei $A=96-156
$ in optimum agreement with the experimental data. The theoretical
spectra was taken to be the optimum one if the excitation energy of the $\ $2%
$^{+}$ state \ $E_{2^{+}}$ was reproduced as closely as possible to the
experimental value. All the parameters were kept fixed throughout the
subsequent calculations. 
The reliability of HFB wave functions
was tested by obtaining an over all agreement between theoretically
calculated results for the yrast spectra, reduced $B(E2$:$0^{+}\rightarrow
2^{+})$ transition probabilities, static quadrupole moments $Q(2^{+})$ as
well as $g$-factors $g(2^{+})$ of the above mentioned nuclei and the
available experimental data. The same PHFB wave functions were employed to
calculate NTMEs $M_{2\nu }$ and half-lives $T_{1/2}^{2\nu }
(0^{+}\rightarrow 0^{+})$ of $\left(e^{+}\beta \beta \right) _{2\nu }$ 
decay for $^{96}$Ru, $^{102}$Pd, $%
^{106,108}$Cd \cite{rain06}, $^{124,126}$Xe, $^{130,132}$Ba \cite{sing07}
and $^{156}$Dy \cite{rath09} isotopes.
It was also shown that the proton-neutron part of the \textit{PQQ}
interaction, which is responsible for triggering deformation in the
intrinsic ground state, plays an important role in the suppression of $%
M_{2\nu }$.

\subsection{Results of $\left( \beta ^{+}\beta ^{+}\right) _{0\nu }$ and $%
\left( \varepsilon \beta ^{+}\right) _{0\nu }$ modes}

The phase space factors $G_{01}$ of $\left( \beta ^{+}\beta ^{+}\right)
_{0\nu }$ and $\left( \varepsilon \beta ^{+}\right) _{0\nu }$ modes have
been evaluated by Doi \textit{\textit{et al.}} with $\ g_{A}=1.261$ \cite
{doi93}. We use the phase space factors after reevaluating them for $%
g_{A}=1.254$. The phase space factors of $\beta ^{+}\beta ^{+}$ $(\varepsilon \beta ^{+})$ 
modes (in yr$^{-1}$) used in the present
calculation are $%
2.243\times 10^{-18}$ $(2.664\times 10^{-17})$, $2.532\times 10^{-18}$ $%
(3.635\times 10^{-17})$, $3.048\times 10^{-18}$ $(5.654\times 10^{-17})$ and 
$5.114\times 10^{-19}$ $(4.901\times 10^{-17})$ for $^{96}$Ru,$^{106}$Cd, $%
^{124}$Xe and $^{130}$Ba nuclei, respectively \cite{doi93}. For $^{102}$Pd
and $^{156}$Dy nuclei, we calculate $G_{01}$ following the notations of Doi 
\textit{\textit{et al. }}\cite{doi93} in the approximation $C_{1}=1.0,$ $%
C_{2}=0.0$, $C_{3}=0.0$ and $R_{1,1}(\varepsilon )=R_{+1}(\varepsilon
)+R_{-1}(\varepsilon )=$1.0. The calculated $G_{01}$ of the $\varepsilon \beta ^{+}$ 
mode for $^{102}$Pd and $^{156}$Dy isotopes are $6.0\times
10^{-19}$ yr$^{-1}$ and 3.250$\times 10^{-17}$ yr$^{-1}$, respectively.

In Table~\ref{tab2}, the NTMEs $M_{F}$, $M_{GT}$, $M_{Fh}$ and $M_{GTh}\,$%
required to study the $\left( \beta ^{+}\beta ^{+}\right) _{0\nu }$ and $\left(
\varepsilon \beta ^{+}\right) _{0\nu }$ modes of $^{96}$Ru, $^{102}$Pd, $%
^{106}$Cd, $^{124}$Xe, $^{130}$Ba and $^{156}$Dy nuclei are compiled.
Following Haxton's prescription \cite{haxt84}, the average energy
denominator is taken as $\overline{A}=1.2A^{1/2}$ MeV. We calculate the four
NTMEs in the approximation of point nucleons, point nucleons plus Jastrow
type of SRC with Miller and Spencer parametrization \cite{mill76}, finite size of nucleons
with dipole form factor and finite size plus SRC. In the case of point
nucleons, the NTMEs $M_{F}$ and $M_{GT}$ are calculated for $\overline{A}$
and $\overline{A}/2$ in the energy denominator$.$ It is observed that the
NTMEs $M_{F}$ and $M_{GT}$ change by 7.8--9.8\% for $\overline{A}/2$ in
comparison to $\overline{A}$ in the energy denominator. Therefore, the
dependence of NTMEs on average excitation energy $\overline{A}$ is small and
the closure approximation is quite good in the case of $\left( \beta
^{+}\beta ^{+}\right) _{0\nu }$ and $\left( \varepsilon \beta ^{+}\right)
_{0\nu }$ modes as expected. In the approximation of light neutrinos, the
NTMEs $M_{F}$ and $M_{GT}$ are reduced by 17.8--21.4\% 
and 12.2--14.2\% for point
nucleon plus SRC, and finite size of nucleons respectively. Finally, the
NTMEs change by 21.7--25.8\% with finite size plus SRC. In the case of heavy
neutrinos, the $M_{Fh}$ and $M_{GTh}$ get reduced by 33.9--38.0\% 
and 65.0--68.5\% with the inclusion of finite size and finite size plus SRC.
\begin{figure}[htb]
\begin{tabular}{c}
\includegraphics [scale=0.388]{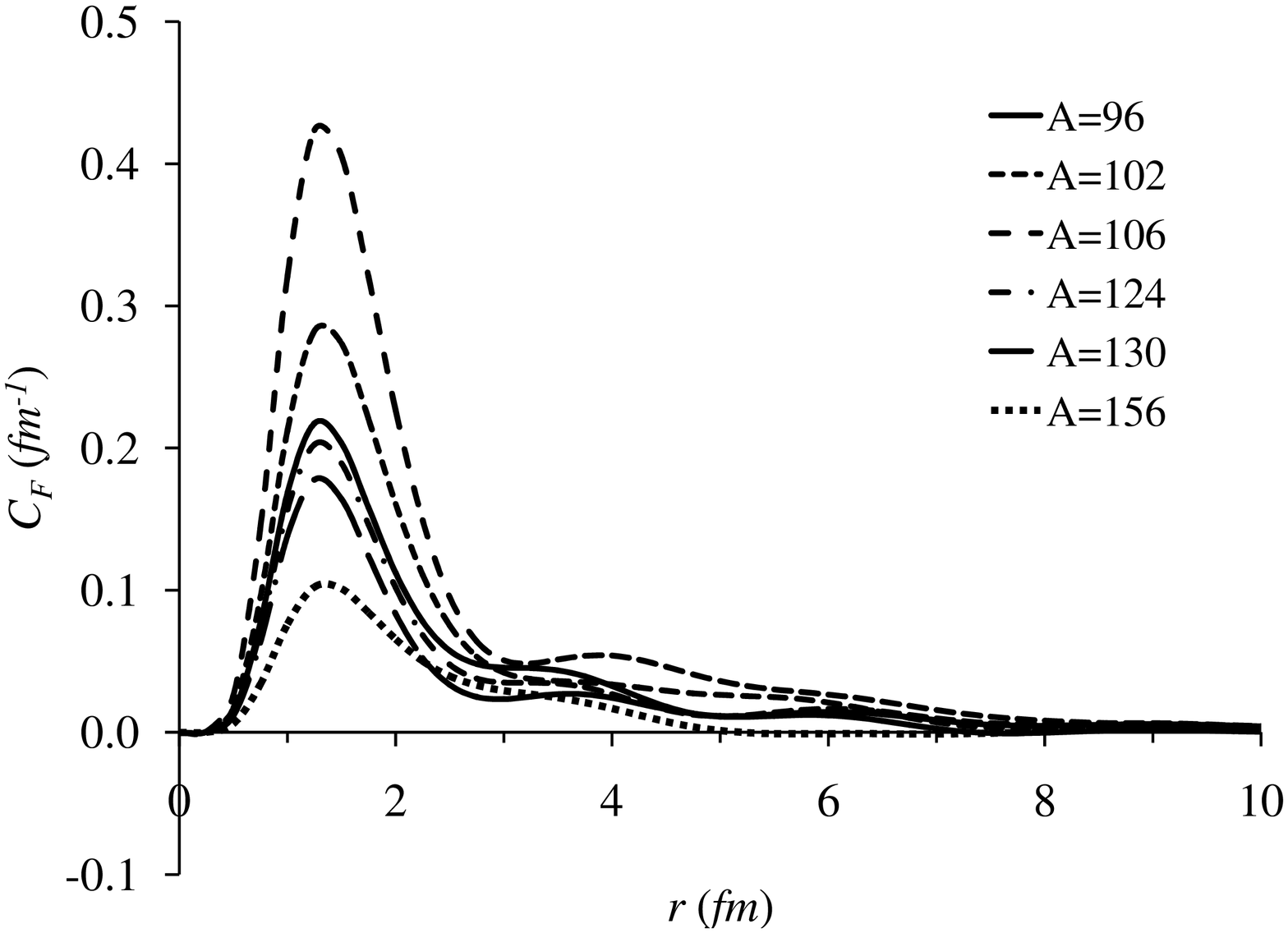} \\
\includegraphics [scale=0.388]{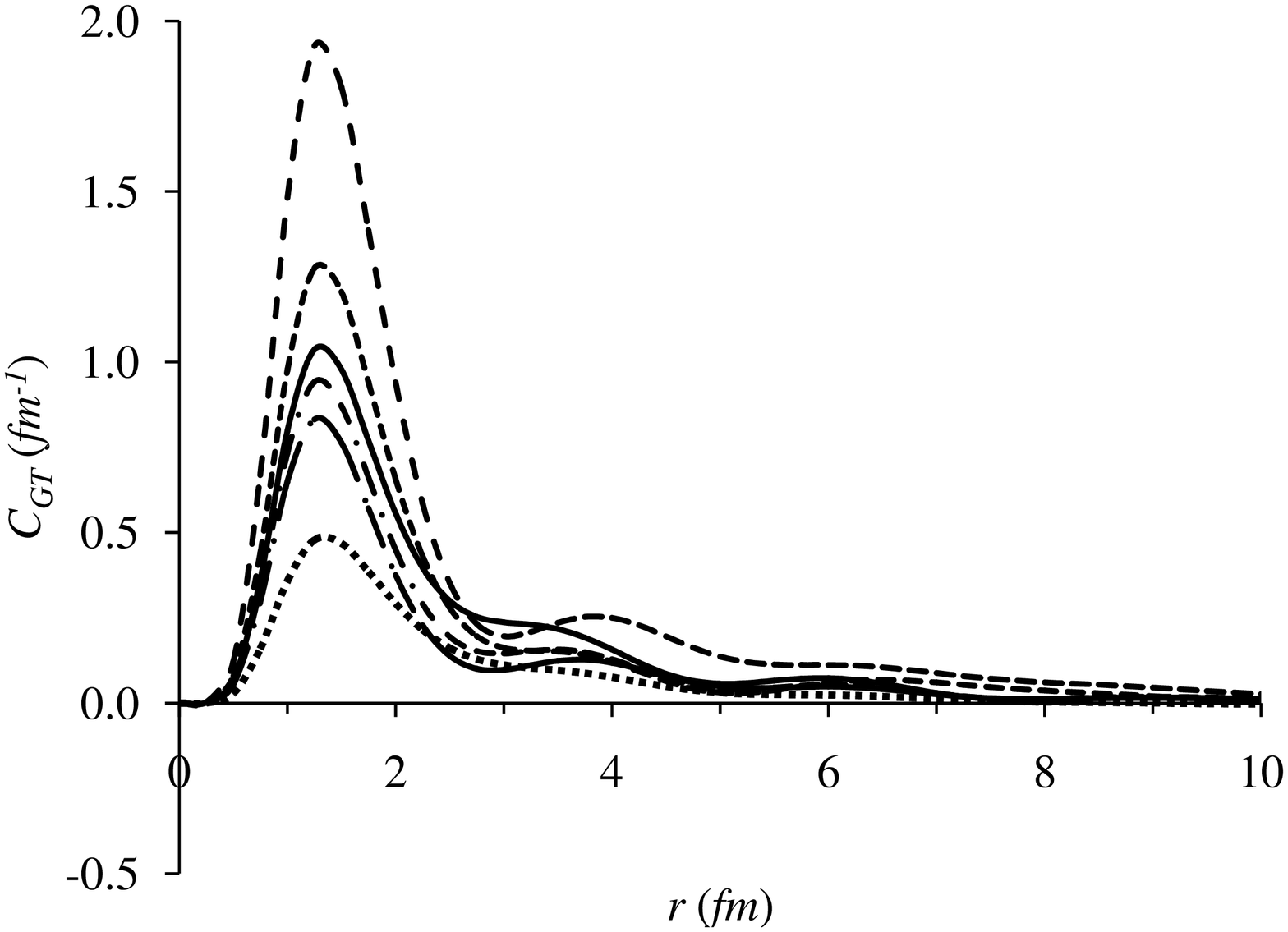} \\
\includegraphics [scale=0.388]{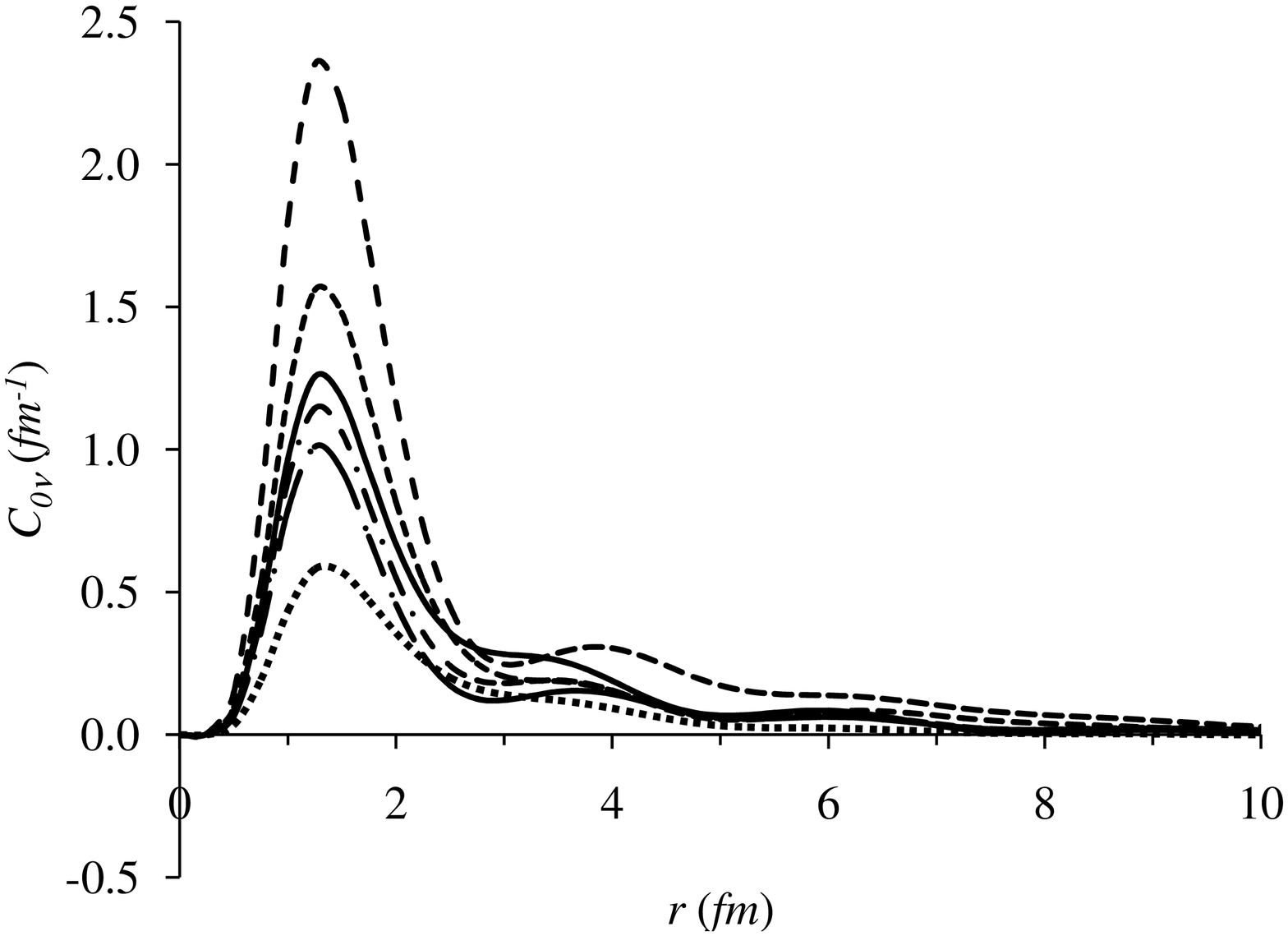} \\
\end{tabular}
\caption{Radial dependence of $C_F(r)$, $C_{GT}(r)$ and $C_{0\nu}(r)$ with FNS
and SRC effects for the $\left( \beta^{+}\beta ^{+}\right) _{0\nu }$ and 
$\left( \varepsilon \beta ^{+}\right)_{0\nu }$ decay modes of $^{96}$Ru, 
$^{102}$Pd, $^{106}$Cd, $^{124}$Xe, $^{130}$Ba and $^{156}$Dy isotopes.}
\label{fig1}
\end{figure}
\begin{table*}[htb]
\caption{Calculated NTMEs for the $0^{+}\rightarrow 0^{+}$ transition 
of $\left( \beta ^{+}\beta ^{+}\right) _{0\nu }$
and $\left( \varepsilon \beta ^{+}\right) _{0\nu }$ modes in the mass mechanism.}
\label{tab2}
\begin{tabular}{ccccrrr}
\hline\hline
Nuclei\thinspace \thinspace & NTMEs & \multicolumn{2}{c}{Point} & \thinspace
\thinspace \thinspace \thinspace \thinspace \thinspace \thinspace Point+SRC
& \thinspace \thinspace \thinspace \thinspace \thinspace \thinspace
\thinspace \thinspace \thinspace \thinspace \thinspace \thinspace \thinspace
\thinspace \thinspace \thinspace \thinspace Extened & \thinspace \thinspace
\thinspace Extended+SRC \\ 
\multicolumn{1}{l}{} & \multicolumn{1}{l}{} & $\,\,\,\,\,\,\,\,\;\,\,\,\,\,%
\overline{A}\,\,\,\,\,\,$ & $\,\,\,\,\,\,\,\,\,\,\,\,\,\,\,\,\,\,\,\overline{%
A}/2\,\,\,\,\,$ & \multicolumn{1}{l}{} & \multicolumn{1}{l}{} & 
\multicolumn{1}{l}{} \\ \hline
\multicolumn{1}{l}{} & \multicolumn{1}{l}{} & \multicolumn{1}{l}{} & 
\multicolumn{1}{l}{} & \multicolumn{1}{l}{} & \multicolumn{1}{l}{} & 
\multicolumn{1}{l}{} \\ 
\multicolumn{1}{l}{$^{96}$Ru} & \multicolumn{1}{l}{$M_{F}$} & 
\multicolumn{1}{r}{0.4983} & \multicolumn{1}{r}{0.5372} & 0.3969 & 0.4309 & 
0.3757 \\ 
\multicolumn{1}{l}{} & \multicolumn{1}{l}{$M_{GT}$} & \multicolumn{1}{r}{
-2.4780} & \multicolumn{1}{r}{-2.6826} & -2.0000 & -2.1591 & -1.8992 \\ 
\multicolumn{1}{l}{} & \multicolumn{1}{l}{$M_{Fh}$} & \multicolumn{1}{r}{
35.8917} & \multicolumn{1}{r}{} & 0 & 22.4117 & 11.4829 \\ 
\multicolumn{1}{l}{} & \multicolumn{1}{l}{$M_{GTh}$} & \multicolumn{1}{r}{
-169.321} & \multicolumn{1}{r}{} & 0 & -106.353 & -54.7130 \\ 
\multicolumn{1}{l}{} & \multicolumn{1}{l}{} & \multicolumn{1}{r}{} & 
\multicolumn{1}{r}{} &  &  &  \\ 
\multicolumn{1}{l}{$^{102}$Pd} & \multicolumn{1}{l}{$M_{F}$} & 
\multicolumn{1}{r}{0.6464} & \multicolumn{1}{r}{0.6995} & 0.5233 & 0.5632 & 
0.4965 \\ 
\multicolumn{1}{l}{} & \multicolumn{1}{l}{$M_{GT}$} & \multicolumn{1}{r}{
-2.7663} & \multicolumn{1}{r}{-2.9861} & -2.1863 & -2.3785 & -2.0631 \\ 
\multicolumn{1}{l}{} & \multicolumn{1}{l}{$M_{Fh}$} & \multicolumn{1}{r}{
43.3140} & \multicolumn{1}{r}{} & 0 & 28.1494 & 14.7508 \\ 
\multicolumn{1}{l}{} & \multicolumn{1}{l}{$M_{GTh}$} & \multicolumn{1}{r}{
-204.336} & \multicolumn{1}{r}{} & 0 & -129.721 & -67.0114 \\ 
\multicolumn{1}{l}{} & \multicolumn{1}{l}{} & \multicolumn{1}{r}{} & 
\multicolumn{1}{r}{} &  &  &  \\ 
\multicolumn{1}{l}{$^{106}$Cd} & \multicolumn{1}{l}{$M_{F}$} & 
\multicolumn{1}{r}{0.9583} & \multicolumn{1}{r}{1.0394} & 0.7704 & 0.8319 & 
0.7299 \\ 
\multicolumn{1}{l}{} & \multicolumn{1}{l}{$M_{GT}$} & \multicolumn{1}{r}{
-4.3495} & \multicolumn{1}{r}{-4.7284} & -3.4635 & -3.7594 & -3.2769 \\ 
\multicolumn{1}{l}{} & \multicolumn{1}{l}{$M_{Fh}$} & \multicolumn{1}{r}{
66.1196} & \multicolumn{1}{r}{} & 0 & 42.5989 & 22.1888 \\ 
\multicolumn{1}{l}{} & \multicolumn{1}{l}{$M_{GTh}$} & \multicolumn{1}{r}{
-311.922} & \multicolumn{1}{r}{} & 0 & -197.061 & -101.408 \\ 
\multicolumn{1}{l}{} & \multicolumn{1}{l}{} & \multicolumn{1}{r}{} & 
\multicolumn{1}{r}{} &  &  &  \\ 
\multicolumn{1}{l}{$^{124}$Xe} & \multicolumn{1}{l}{$M_{F}$} & 
\multicolumn{1}{r}{0.4865} & \multicolumn{1}{r}{0.5333} & 0.3915 & 0.4233 & 
0.3717 \\ 
\multicolumn{1}{l}{} & \multicolumn{1}{l}{$M_{GT}$} & \multicolumn{1}{r}{
-2.1387} & \multicolumn{1}{r}{-2.3299} & -1.6905 & -1.8416 & -1.5978 \\ 
\multicolumn{1}{l}{} & \multicolumn{1}{l}{$M_{Fh}$} & \multicolumn{1}{r}{
33.7569} & \multicolumn{1}{r}{} & 0 & 21.1455 & 10.8449 \\ 
\multicolumn{1}{l}{} & \multicolumn{1}{l}{$M_{GTh}$} & \multicolumn{1}{r}{
-159.250} & \multicolumn{1}{r}{} & 0 & -98.9817 & -50.4944 \\ 
\multicolumn{1}{l}{} & \multicolumn{1}{l}{} & \multicolumn{1}{r}{} & 
\multicolumn{1}{r}{} &  &  &  \\ 
\multicolumn{1}{l}{$^{130}$Ba} & \multicolumn{1}{l}{$M_{F}$} & 
\multicolumn{1}{r}{0.4183} & \multicolumn{1}{r}{0.4593} & 0.3338 & 0.3623 & 
0.3163 \\ 
\multicolumn{1}{l}{} & \multicolumn{1}{l}{$M_{GT}$} & \multicolumn{1}{r}{
-1.8626} & \multicolumn{1}{r}{-2.0325} & -1.4633 & -1.5986 & -1.3812 \\ 
\multicolumn{1}{l}{} & \multicolumn{1}{l}{$M_{Fh}$} & \multicolumn{1}{r}{
30.0461} & \multicolumn{1}{r}{} & 0 & 18.7025 & 9.5438 \\ 
\multicolumn{1}{l}{} & \multicolumn{1}{l}{$M_{GTh}$} & \multicolumn{1}{r}{
-141.744} & \multicolumn{1}{r}{} & 0 & -87.8418 & -44.6828 \\ 
\multicolumn{1}{l}{} & \multicolumn{1}{l}{} & \multicolumn{1}{r}{} & 
\multicolumn{1}{r}{} &  &  &  \\ 
\multicolumn{1}{l}{$^{156}$Dy} & \multicolumn{1}{l}{$M_{F}$} & 
\multicolumn{1}{r}{0.2461} & \multicolumn{1}{r}{0.2698} & 0.2022 & 0.2160 & 
0.1926 \\ 
\multicolumn{1}{l}{} & \multicolumn{1}{l}{$M_{GT}$} & \multicolumn{1}{r}{
-1.1281} & \multicolumn{1}{r}{-1.2319} & -0.9208 & -0.9867 & -0.8754 \\ 
\multicolumn{1}{l}{} & \multicolumn{1}{l}{$M_{Fh}$} & \multicolumn{1}{r}{
15.7014} & \multicolumn{1}{r}{} & 0 & 10.3729 & 5.4997 \\ 
\multicolumn{1}{l}{} & \multicolumn{1}{l}{$M_{GTh}$} & \multicolumn{1}{r}{
-74.0722} & \multicolumn{1}{r}{} & 0 & -48.6696 & -25.6980 \\ 
\multicolumn{1}{l}{} & \multicolumn{1}{l}{} & \multicolumn{1}{r}{} & 
\multicolumn{1}{r}{} &  &  &  \\ \hline\hline
\end{tabular}
\end{table*}

The radial dependence of $C_{0\nu}(r)$ defined by
\begin{equation}
M_{0\nu }=\int\limits_{0}^{\infty }C_{0\nu}(r)\;dr
\end{equation}
has been studied in the QRPA by \v{S}imkovic 
{\it et al.} \cite{simk08} and ISM by Men\'{e}ndez {\it et al.} \cite{mene09}.
In both QRPA and ISM calculations, it has been established that the contributions 
of decaying pairs coupled to $J=0$ and $J > 0$ almost cancel 
beyond $r \approx 3$ fm and the magnitude of $C_{0\nu}(r)$ for all nuclei 
undergoing $\left( \beta^{-}\beta ^{-}\right) _{0\nu }$ decay are the maximum 
about the internucleon distance $r \approx 1$ fm. In Fig.~\ref{fig1}, we plot the radial dependence
of the total matrix elements $C_{0\nu}(r)$ as well as their Fermi and Gamow-Teller 
components due to the 
exchange of light neutrinos. It is noticed that the maximum value of 
$C_F(r)$, $C_{GT}(r)$ and $C_{0\nu}(r)$ 
is at $r=1.25$ fm in agreement with the works done by  \v{S}imkovic
{\it et al.} \cite{simk08} and Men\'{e}ndez {\it et al.} \cite{mene09}.

In Table~\ref{tab3}, we tabulate the extracted limits on the effective light
neutrino mass $<m_{\nu }>$ as well as heavy neutrino mass $<M_{N}>$ using
presently available experimentally observed limits on half-lives of $\left(
\beta ^{+}\beta ^{+}\right) _{0\nu }$ and $\left( \varepsilon \beta
^{+}\right) _{0\nu }$ modes. It is observed that limits on $<m_{\nu }>$
\begin{table*}[htb]
\caption{Upper and lower bounds on light and heavy neutrino effective
masses $<m_{\nu }>$ and $<M_{N}>$, respectively, for the $\left( \beta
^{+}\beta ^{+}\right) _{0\nu }$ and $\left( \varepsilon \beta ^{+}\right)
_{0\nu }$ modes of $^{96}$Ru, $^{106}$Cd, $^{124}$Xe and $^{130}$Ba isotopes.}
\label{tab3}
\begin{tabular}{cccccccl}
\hline\hline
Nuclei\thinspace \thinspace \thinspace \thinspace \thinspace \thinspace  & 
\multicolumn{2}{c}{$T_{1/2}^{0\nu }$(yr)} & Ref.\thinspace \thinspace
\thinspace \thinspace \thinspace \thinspace \thinspace \thinspace \thinspace
\thinspace  & \multicolumn{2}{c}{$<m_{\nu }>(eV)$}~~~~~~~ & \multicolumn{2}{c}{$%
<M_{N}>(GeV)$} \\ 
\multicolumn{1}{l}{} & ~~~~$\beta ^{+}\beta
^{+}\,\,\,\,\,\,\,\,\,\,\,\,\,\,\,\,\,\,\,\,\,\,\,\,\,\,\,\,\,$ & ~~~~$%
\varepsilon \beta ^{+}\,\,\,\,\,\,\,\,\,\,\,\,\,\,\,\,\,\,\,\,\,\,\,\,$ &  & 
~~~~$\beta ^{+}\beta ^{+}\,\,\,\,\,\,\,\,\,\,\,\,\,\,\,\,\,\,\,\,\,\,$ & ~~~~~$%
\varepsilon \beta
^{+}\,\,\,\,\,\,\,\,\,\,\,\,\,\,\,\,\,\,\,\,\,\,\,\,\,\,\,\,\,$ &~~$\beta
^{+}\beta ^{+}\,\,\,\,\,\,\,\,\,\,\,\,\,\,\,\,\,\,\,\,\,\,\,$ &~~~~$\varepsilon
\beta ^{+}$ \\ \hline
\multicolumn{1}{l}{$^{96}$Ru} & \multicolumn{1}{l}{$>3.1\times 10^{16}$} & 
\multicolumn{1}{l}{$>6.7\times 10^{16}$} & \multicolumn{1}{l}{\cite{norm85}}
& \multicolumn{1}{l}{$8.52\times 10^{5}$} & \multicolumn{1}{l}{$\text{ }%
1.68\times 10^{5}$} & \multicolumn{1}{l}{$16.38$} & $82.98$ \\ 
\multicolumn{1}{l}{$^{106}$Cd} & \multicolumn{1}{l}{$>1.4\times 10^{19}$} & 
\multicolumn{1}{l}{$>7.0\times 10^{19}$} & \multicolumn{1}{l}{\cite{dane03}}
& \multicolumn{1}{l}{$2.14\times 10^{4}$} & \multicolumn{1}{l}{$\text{ }%
2.53\times 10^{3}$} & \multicolumn{1}{l}{$6.90\times 10^{2}$} & $5.85\times
10^{3}$ \\ 
\multicolumn{1}{l}{$^{124}$Xe} & \multicolumn{1}{l}{$>4.2\times 10^{17}$} & 
\multicolumn{1}{l}{$>1.2\times 10^{18}$} & \multicolumn{1}{l}{\cite{bara89}}
& \multicolumn{1}{l}{$2.29\times 10^{5}$} & \multicolumn{1}{l}{$\text{ }%
3.15\times 10^{4}$} & \multicolumn{1}{l}{$65.12$} & $4.74\times 10^{2}$ \\ 
\multicolumn{1}{l}{$^{130}$Ba} & \multicolumn{1}{l}{$>4.0\times 10^{21}$} & 
\multicolumn{1}{l}{$>4.0\times 10^{21}$} & \multicolumn{1}{l}{\cite{bara96a}}
& \multicolumn{1}{l}{$6.66\times 10^{3}$} & \multicolumn{1}{l}{$\text{ }%
6.80\times 10^{2}$} & \multicolumn{1}{l}{$2.30\times 10^{3}$} & $2.25\times
10^{4}$ \\ \hline\hline
\end{tabular}
\end{table*}
and $<M_{N}>$ are not so much stringent as in the case of $\left( \beta
^{-}\beta ^{-}\right) _{0\nu }$ mode. Further, better limits are
obtained in the case of $\left( \varepsilon \beta ^{+}\right) _{0\nu }$ mode
even for equal limits on half-lives of $\left( \beta ^{+}\beta ^{+}\right)
_{0\nu }$ and $\left( \varepsilon \beta ^{+}\right) _{0\nu }$ modes.
In the case of $\left( \varepsilon \beta^{+}\right) _{0\nu }$ mode, 
the best limits obtained for $^{130}$Ba nuclei are $<m_{\nu}>\,<6.8\times 10^{2}$ eV and
$<M_{N}>\, >2.25\times 10^{4}$ GeV.

In Table~\ref{tab4}, we compile available theoretical results 
in other nuclear models along
with ours. To the best of our knowledge, no theoretical result and
\begin{table*}[htb]
\caption{Predicted half-lives $T_{1/2}^{0\nu }$ $<m_{\nu }>^{2}$
of $\left( \beta ^{+}\beta ^{+}\right) _{0\nu }$ and $\left( \varepsilon
\beta ^{+}\right) _{0\nu }$ modes due to the exchange of light neutrino and
extracted limits on effective heavy neutrino mass $\left\langle
M_{N}\right\rangle $ from the same predicted half-lives for $<m_{\nu }>=1$
eV. The $\dagger $ and $\ddagger $ denote WS and AWS basis respectively in
reference \cite{auno98}.} 
\label{tab4}
\begin{tabular}{lllrrrllllllc}
\hline\hline
{\small Nuclei}~~~~ & {\small Model} & {\small Ref.} & $M_{F}$ & $M_{GT}$ & $%
\left| M_{0\nu }\right| $ & \thinspace \thinspace \thinspace \thinspace
\thinspace \thinspace \thinspace \thinspace \thinspace  & \multicolumn{2}{c}{%
$T_{1/2}^{0\nu }$ $<m_{\nu }>^{2}$~~~~} & $M_{Fh}$ & $M_{GTh}$ & $\left|
M_{0N}\right| $ & $<M_{N}>$ \\ 
&  &  &  &  &  &  & \multicolumn{2}{c}{(yr eV$^{2}$)} &  &  &  & 
\multicolumn{1}{l}{\thinspace \thinspace \thinspace \thinspace \thinspace
\thinspace {\small (GeV)}} \\ 
&  &  &  &  &  &  & ~~~~$\beta ^{+}\beta ^{+}$ & ~~~~~$\varepsilon \beta ^{+}
$ &  &  &  & \multicolumn{1}{l}{} \\ \hline
&  &  & \thinspace \thinspace \thinspace \thinspace \thinspace \thinspace
\thinspace \thinspace \thinspace \thinspace \thinspace \thinspace \thinspace
\thinspace \thinspace \thinspace \thinspace \thinspace \thinspace \thinspace
\thinspace  & \thinspace \thinspace \thinspace \thinspace \thinspace
\thinspace \thinspace \thinspace \thinspace \thinspace \thinspace \thinspace
\thinspace \thinspace \thinspace \thinspace \thinspace \thinspace \thinspace
\thinspace \thinspace  & \thinspace \thinspace \thinspace \thinspace
\thinspace \thinspace \thinspace \thinspace \thinspace \thinspace \thinspace
\thinspace \thinspace \thinspace \thinspace \thinspace \thinspace \thinspace
\thinspace \thinspace \thinspace \thinspace \thinspace \thinspace  &  & 
\thinspace \thinspace \thinspace \thinspace \thinspace \thinspace \thinspace
\thinspace \thinspace \thinspace \thinspace \thinspace \thinspace \thinspace
\thinspace \thinspace \thinspace \thinspace \thinspace \thinspace \thinspace
\thinspace \thinspace \thinspace \thinspace \thinspace \thinspace \thinspace
\thinspace \thinspace \thinspace \thinspace \thinspace \thinspace \thinspace
\thinspace \thinspace  & \thinspace \thinspace \thinspace \thinspace
\thinspace \thinspace \thinspace \thinspace \thinspace \thinspace \thinspace
\thinspace \thinspace \thinspace \thinspace \thinspace \thinspace \thinspace
\thinspace \thinspace \thinspace \thinspace \thinspace \thinspace \thinspace
\thinspace \thinspace \thinspace \thinspace \thinspace \thinspace \thinspace
\thinspace \thinspace \thinspace \thinspace \thinspace  & \thinspace
\thinspace \thinspace \thinspace \thinspace \thinspace \thinspace \thinspace
\thinspace \thinspace \thinspace \thinspace \thinspace \thinspace \thinspace
\thinspace \thinspace \thinspace \thinspace \thinspace \thinspace \thinspace 
& \thinspace \thinspace \thinspace \thinspace \thinspace \thinspace
\thinspace \thinspace \thinspace \thinspace \thinspace \thinspace \thinspace
\thinspace \thinspace \thinspace \thinspace \thinspace \thinspace \thinspace
\thinspace \thinspace \thinspace \thinspace  & \thinspace \thinspace
\thinspace \thinspace \thinspace \thinspace \thinspace \thinspace \thinspace
\thinspace \thinspace \thinspace \thinspace \thinspace \thinspace \thinspace
\thinspace \thinspace \thinspace \thinspace \thinspace \thinspace \thinspace
\thinspace \thinspace  & \multicolumn{1}{l}{} \\ 
$^{96}${\small \ Ru} & {\small PHFB} &  & {\small 0.376} & {\small -1.899}
& {\small 2.275} &  & {\small \allowbreak 2.\thinspace 249}$\times ${\small %
10}$^{28}$ & {\small \allowbreak 1.\thinspace 894}$\times ${\small 10}$^{27}$
& {\small 11.483} & {\small -54.713} & {\small 66.196} & \multicolumn{1}{l}{%
{\small 1.\thinspace 40}$\times ${\small 10}$^{7}$} \\ 
& {\small MCM} & {\small \cite{suho03}} & {\small -0.705} & {\small 1.678} & 
{\small 2.383} &  & {\small 2.\thinspace 050}$\times ${\small 10}$^{28}$ & 
{\small 1.\thinspace 726}$\times ${\small 10}$^{27}$ &  &  &  & 
\multicolumn{1}{l}{} \\ 
& {\small QRPA} & {\small \cite{hirs94}} & {\small -0.98} & {\small 2.62} & 
{\small 3.60} &  & {\small \allowbreak 8.\thinspace 981}$\times ${\small 10}$%
^{27}$ & {\small \allowbreak 7.563}$\times ${\small 10}$^{26}$ &  &  &  & 
\multicolumn{1}{l}{} \\ 
& {\small QRPA} & {\small \cite{stau91}} &  &  & {\small 4.228} &  & {\small %
\allowbreak 6.\thinspace 511}$\times ${\small 10}$^{27}$ & {\small %
\allowbreak 5.\thinspace 483}$\times ${\small 10}$^{26}$ &  &  &  & 
\multicolumn{1}{l}{} \\ 
&  &  &  &  &  &  &  &  &  &  &  &  \\ 
$^{102}${\small \ Pd} &  &  & {\small 0.497} & {\small -2.063} & {\small %
2.560} &  &  & {\small 6.643}$\times ${\small 10}$^{28}$ & {\small 14.751} & 
{\small -67.011} & {\small 81.762} & {\small 1.53}$\times ${\small 10}$^{7}$
\\ 
&  &  &  &  &  &  &  &  &  &  &  & \multicolumn{1}{l}{} \\ 
$^{106}${\small \ Cd} & {\small PHFB} &  & {\small 0.730} & {\small -3.277}
& {\small 4.007} &  & {\small \allowbreak 6.\thinspace 424}$\times ${\small %
10}$^{27}$ & {\small \allowbreak 4.\thinspace 474}$\times ${\small 10}$^{26}$
& {\small 22.189} & {\small -101.408} & {\small 123.597} & 
\multicolumn{1}{l}{{\small \allowbreak 1.\thinspace 48}$\times ${\small 10}$%
^{7}$} \\ 
& {\small MCM} & {\small \cite{suho03}} & {\small -1.191} & {\small 2.203} & 
{\small 3.394} &  & {\small \allowbreak 8.\thinspace 953}$\times ${\small 10}%
$^{27}$ & {\small 6.\thinspace 236}$\times ${\small 10}$^{26}$ &  &  &  & 
\multicolumn{1}{l}{} \\ 
& {\small SQRPA(l)} & {\small \cite{stoi03}} & {\small -2.12} & {\small 5.73}
& {\small 7.85} &  & {\small \allowbreak 1.674}$\times ${\small 10}$^{27}$ & 
{\small \allowbreak 1.\thinspace 166}$\times ${\small 10}$^{26}$ &  &  &  & 
\multicolumn{1}{l}{} \\ 
& {\small SQRPA(s)} & {\small \cite{stoi03}} & {\small -2.18} & {\small 5.99}
& {\small 8.17} &  & {\small 1.\thinspace 545}$\times ${\small 10}$^{27}$ & 
{\small 1.\thinspace 076}$\times ${\small 10}$^{26}$ &  &  &  & 
\multicolumn{1}{l}{} \\ 
& {\small QRPA} & {\small \cite{hirs94}} & {\small -1.22} & {\small 3.34} & 
{\small 4.56} &  & {\small \allowbreak 4.\thinspace 960}$\times ${\small 10}$%
^{27}$ & {\small \allowbreak 3.\thinspace 455}$\times ${\small 10}$^{26}$ & 
&  &  & \multicolumn{1}{l}{} \\ 
& {\small QRPA} & {\small \cite{stau91}} &  &  & {\small 4.778} &  & {\small %
\allowbreak 4.\thinspace 517}$\times ${\small 10}$^{27}$ & {\small %
\allowbreak \allowbreak 3.\thinspace 146}$\times ${\small 10}$^{26}$ &  &  & 
& \multicolumn{1}{l}{} \\ 
&  &  &  &  &  &  &  &  &  &  &  & \multicolumn{1}{l}{} \\ 
$^{124}${\small \ Xe} & {\small PHFB} &  & {\small 0.372} & {\small -1.598}
& {\small 1.970} &  & {\small \allowbreak 2.\thinspace 208}$\times ${\small %
10}$^{28}$ & {\small \allowbreak 1.\thinspace 191}$\times ${\small 10}$^{27}$
& {\small 10.845} & {\small -50.494} & {\small 61.339} & \multicolumn{1}{l}{%
{\small 1.\thinspace 49}$\times ${\small 10}$^{7}$} \\ 
& {\small MCM} & {\small \cite{suho03}} & {\small -2.572} & {\small 5.729} & 
{\small 8.301} &  & {\small 1.\thinspace 243}$\times ${\small 10}$^{27}$ & 
{\small \allowbreak 6.\thinspace 703}$\times ${\small 10}$^{25}$ &  &  &  & 
\multicolumn{1}{l}{} \\ 
& {\small QRPA}$^{\dagger }$ & {\small \cite{auno98}} & {\small -2.236} & 
{\small 5.128} & {\small 7.364} &  & {\small \allowbreak \allowbreak
1.\thinspace 580}$\times ${\small 10}$^{27}$ & {\small \allowbreak
\allowbreak 8.\thinspace 517}$\times ${\small 10}$^{25}$ &  &  &  & 
\multicolumn{1}{l}{} \\ 
& {\small QRPA}$^{\ddagger }$ & {\small \cite{auno98}} & {\small -2.574} & 
{\small 5.733} & {\small 8.307} &  & {\small \allowbreak 1.\thinspace 241}$%
\times ${\small 10}$^{27}$ & {\small \allowbreak 6.\thinspace 693}$\times $%
{\small 10}$^{25}$ &  &  &  & \multicolumn{1}{l}{} \\ 
& {\small QRPA} & {\small \cite{hirs94}} & {\small -1.35} & {\small 3.92} & 
{\small 5.27} &  & {\small \allowbreak 3.\thinspace 084}$\times ${\small 10}$%
^{27}$ & {\small \allowbreak 1.\thinspace 663}$\times ${\small 10}$^{26}$ & 
&  &  & \multicolumn{1}{l}{} \\ 
& {\small QRPA} & {\small \cite{stau91}} &  &  & {\small 2.975} &  & {\small %
\allowbreak 9.\thinspace 678}$\times ${\small 10}$^{27}$ & {\small %
\allowbreak \allowbreak 5.\thinspace 218}$\times ${\small 10}$^{26}$ &  &  & 
& \multicolumn{1}{l}{} \\ 
&  &  &  &  &  &  &  &  &  &  &  & \multicolumn{1}{l}{} \\ 
$^{130}${\small Ba} & {\small PHFB} &  & {\small 0.316} & {\small -1.381} & 
{\small 1.697} &  & {\small \allowbreak 1.\thinspace 772}$\times ${\small 10}%
$^{29}$ & {\small \allowbreak 1.\thinspace 849}$\times ${\small 10}$^{27}$ & 
{\small 9.544} & {\small -44.683} & {\small 54.227} & \multicolumn{1}{l}{%
{\small 1.\thinspace 53}$\times ${\small 10}$^{7}$} \\ 
& {\small MCM} & {\small \cite{suho03}} & {\small -1.748} & {\small 3.382} & 
{\small 5.130} &  & {\small \allowbreak \allowbreak 1.\thinspace 940}$\times 
${\small 10}$^{28}$ & {\small \allowbreak 2.\thinspace 025}$\times ${\small %
10}$^{26}$ &  &  &  & \multicolumn{1}{l}{} \\ 
& {\small QRPA} & {\small \cite{hirs94}} & {\small -1.50} & {\small 4.02} & 
{\small 5.52} &  & {\small \allowbreak 1.\thinspace 676}$\times ${\small 10}$%
^{28}$ & {\small 1.749}$\times ${\small 10}$^{26}$ &  &  &  & 
\multicolumn{1}{l}{} \\ 
& {\small QRPA} & {\small \cite{stau91}} &  &  & {\small 5.579} &  & {\small %
\allowbreak 1.\thinspace 641}$\times ${\small 10}$^{28}$ & {\small %
1.\thinspace 712}$\times ${\small 10}$^{26}$ &  &  &  & \multicolumn{1}{l}{}
\\ 
&  &  &  &  &  &  &  &  &  &  &  &  \\ 
$^{156}${\small Dy} & {\small PHFB} &  & {\small 0.193} & {\small -0.875} & 
{\small 1.068} &  &  & {\small 7.044}$\times ${\small 10}$^{27}$ & {\small %
5.500} & {\small -25.698} & {\small 31.198} & {\small 1.\thinspace 40}$%
\times ${\small 10}$^{7}$ \\ 
&  &  &  &  &  &  &  &  &  &  &  & \multicolumn{1}{l}{} \\ \hline\hline
\end{tabular}
\end{table*}
experimental half-life limit is available for $^{102}$Pd and $^{156}$Dy
isotopes. Staudt \textit{et al.} \cite{stau91} have reported only NTMEs $%
\left| M_{0\nu }\right| =\left| M_{GT}-M_{F}\right| $ in the mass mechanism.
In the QRPA calculations of Hirsch \textit{et al. }\cite{hirs94} and Staudt 
\textit{et al.} \cite{stau91}, 
the former used two major oscillator shells, where as 
the latter used a model space consisting of $%
3\hbar \omega +4\hbar \omega +0h_{9/2}+0h_{11/2}$ orbits.
The used SPEs are identical. Both
the calculation use a realistic effective two body interaction using Paris
potential. The NTMEs $\left| M_{0\nu }\right| $ are almost identical in both
the QRPA calculations but for $^{124}$Xe, where a difference by
a factor of 1.8 approximately is noticed. In the SQRPA model, Stoica \textit{et al.} 
\cite{stoi03} have studied 
$\left( \beta ^{+}\beta ^{+}\right) _{0\nu }$ and $\left(
\varepsilon \beta ^{+}\right) _{0\nu }$ modes of $^{106}$Cd isotope
using two model spaces, namely small basis (oscillator
shells of $3\hbar \omega -5\hbar \omega +i_{13/2}$ orbits) and a large basis
(oscillator shells of $2\hbar \omega -5\hbar \omega +i_{13/2}$ orbits) with
two-body effective interactions derived
from the Bonn-A potential. The NTMEs calculated in the SQRPA \cite{stoi03}
do not depend much on the model space and differ by a factor of 1.8
approximately from those of Hirsch \textit{et al.} \cite{hirs94}. In the MCM, 
Suhonen \textit {et al.} \cite{suho03} have studied the 
$\left( \beta ^{+}\beta ^{+}\right) _{0\nu }$ and $\left(
\varepsilon \beta ^{+}\right) _{0\nu }$ modes of $^{96}$Ru, $^{106}$Cd,
$^{124}$Xe and $^{130}$Ba nuclei. 
It is worth mentioning that besides the model space, SPEs and effective 
two-body interaction, different values of $g_A$,
specifically $g_A=1.254$ \cite{hirs94,stau91,stoi03} 
and $1.0$ \cite{suho03,auno98},
are also used in these calulations. 

The calculated NTMEs $\left| M_{0\nu }\right| $ in the PHFB model for the $^{96}$Ru and $^{106}$Cd isotopes
are very close to those obtanied in the MCM, and in the later case also to the QRPA results.
For $^{124}$Xe and $^{130}$Ba isotopes, the NTMEs
are smaller than those in other models and
this is reflected in half-lives which are up to one order of magnitude longer.
As the extracted limits on the effective neutrino masses $<m_{\nu }>$
and $<M_{N}>$ are not stringent enough, it is more meaningful to calculate
half-lives of $\left( \beta ^{+}\beta ^{+}\right) _{0\nu }$ and $\left(
\varepsilon \beta ^{+}\right) _{0\nu }$ modes, which will be useful for the
design of future experimental set ups. Hence, we calculate half-lives of $%
\left( \beta ^{+}\beta ^{+}\right) _{0\nu }$ and $\left( \varepsilon \beta
^{+}\right) _{0\nu }$ modes for $<m_{\nu }>=1$ $eV$ and extract
corresponding limits on heavy neutrino mass $<M_{N}>$, which are given in
the same Table~\ref{tab4}.

In the mass mechanisms, there are two noteworthy observations. The equality
in NTMEs of $\left( \beta ^{+}\beta ^{+}\right) _{0\nu }$ and $\left(
\varepsilon \beta ^{+}\right) _{0\nu }$ modes implies that 
\begin{equation}
\frac{T_{1/2}^{0\nu }\left( \beta ^{+}\beta ^{+}\right) }{T_{1/2}^{0\nu
}\left( \varepsilon \beta ^{+}\right) }=\frac{G_{01}\left( \varepsilon \beta
^{+}\right) }{G_{01}\left( \beta ^{+}\beta ^{+}\right) }.
\end{equation}
Therefore, the experimental observation of $\left( \varepsilon \beta
^{+}\right) _{0\nu }$ mode will provide the half-life $T_{1/2}^{0\nu }\left(
\beta ^{+}\beta ^{+}\right) $ of $\left( \beta ^{+}\beta ^{+}\right) _{0\nu }
$ mode as the phase space factors are exactly calculable. Further, it is
noticed that the ratios of $\left| M_{0\nu }\right| $ and $\left|
M_{0N}\right| $ given in Table~\ref{tab2} are almost constant for different nuclei
and $\left| M_{0N}\right| /\left| M_{0\nu }\right| \approx 29-32$
approximately. Similar behaviour of the ratios $\left| M_{0N}\right| /\left|
M_{0\nu }\right| $ $\approx 28-30$ is also observed for the NTMEs of $%
\left( \beta ^{-}\beta ^{-}\right) _{0\nu }$ mode \cite{chat08}. This
implies that in the mass mechanism, the half-lives for different nuclei due
to exchange of light and heavy neutrinos are also in constant ratio 
\begin{equation}
\frac{T_{1/2}^{0\nu }(m_{\nu })}{T_{1/2}^{0\nu }(M_{N})}\propto \frac{\left|
M_{0N}\right| ^{2}}{\left| M_{0\nu }\right| ^{2}}.
\end{equation}
It will be interesting to verify whether the observed constancy of 
$\left| M_{0N}\right| /\left|M_{0\nu }\right|$ in different nuclei is a 
generic feature or artifact of the present calculation.

\subsection{Quadrupolar correlations and deformation effects}

As already mentioned, the quadrupolar correlations are mainly responsible
for the deformation of nuclei. To understand the role of deformation on
NTMEs $M_{\alpha }$ $(\alpha =F,GT,Fh,GTh)$ of $\left( \beta ^{+}\beta
^{+}\right) _{0\nu }$ and $\left( \varepsilon \beta ^{+}\right) _{0\nu }$ modes,
we investigate the variation of the latter by changing the strength of the 
\textit{QQ} interaction $\zeta _{qq}$ for the case in which NTMEs are
calculated with finite size and short range correlations. It is
observed that in general, there is an inverse correlation between the
magnitudes of NTMEs and quadrupole moments $Q(2^{+})$ as well as deformation parameters 
$\beta _{2}$. Further, the effect of deformation on $M_{\alpha }$ is
quantified by defining a quantity $D_{\alpha }$ as the ratio of $M_{\alpha }$
at zero deformation ($\zeta _{qq}=0$) and full deformation ($\zeta _{qq}=1$%
). The $D_{\alpha }$ is given by 
\begin{equation}
D_{\alpha }=\frac{M_{\alpha }(\zeta _{qq}=0)}{M_{\alpha }(\zeta _{qq}=1)}.
\end{equation}
The tabulated values of $D_{\alpha }$ in Table~\ref{tab5} for $^{96}$Ru, $^{102}$Pd, $%
^{106}$Cd, $^{124}$Xe, $^{130}$Ba and $^{156}$Dy nuclei suggest that the
NTMEs $M_{\alpha }$ are
suppressed by factor of 1.7--10.7 in the mass range $A=96-156$ due to
deformation effects. We also give the same deformation ratio $D_{2\nu }$ for
comparison in the last row of the same table, which also change by almost
same amount due to the deformation effects. Hence, it is clear that the
deformation effects are important for $\left( \beta ^{+}\beta
^{+}\right) _{0\nu }$ and $\left( \varepsilon \beta ^{+}\right) _{0\nu }$
modes as well as $\left( e^{+}\beta \beta \right) _{2\nu }$ decay so far as
the nuclear structure aspect of $e^{+}\beta \beta $ decay is concerned.

\begin{figure*}[htb]
\begin{tabular}{cc}
\includegraphics [scale=0.40]{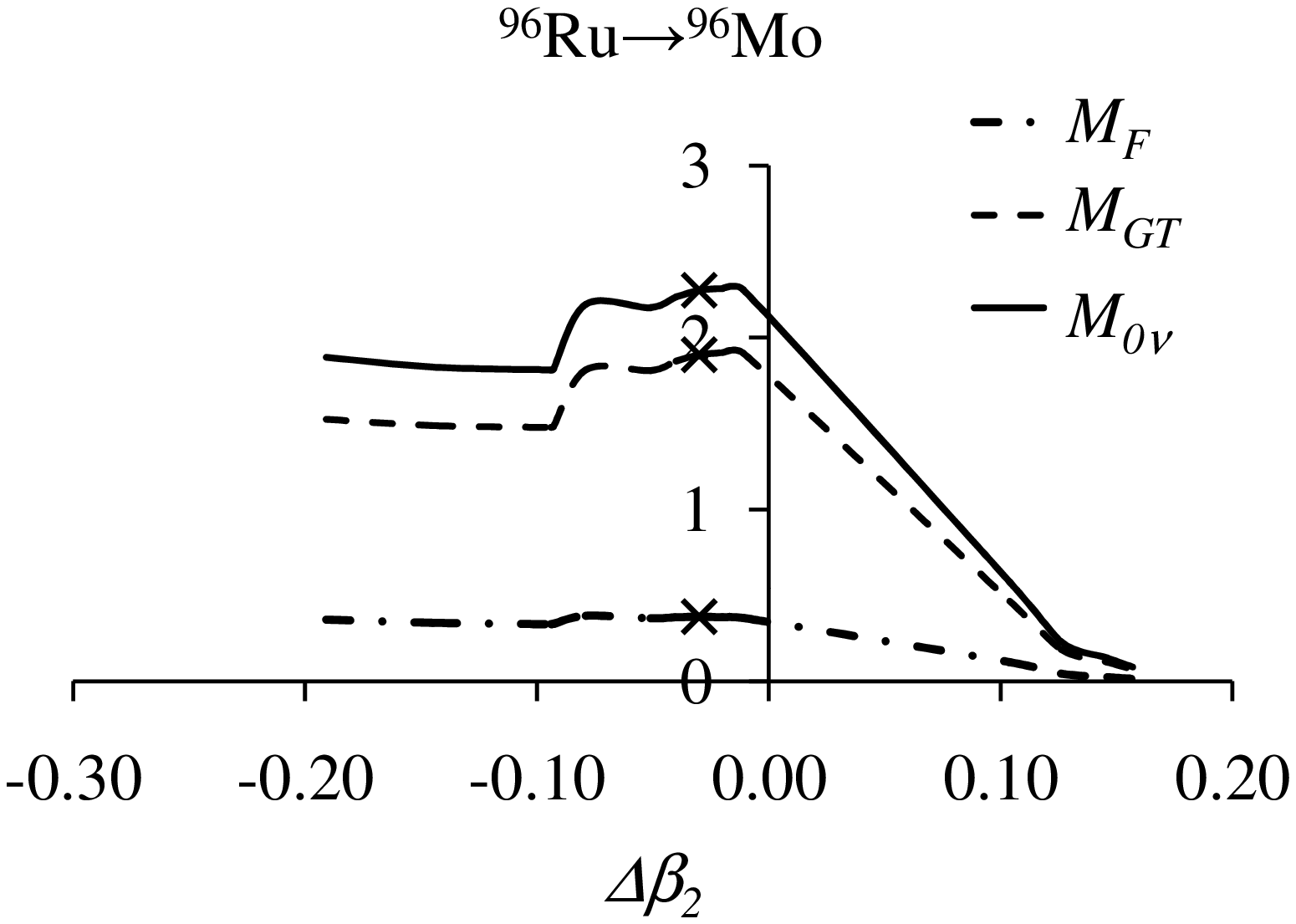} & \includegraphics [scale=0.40]{%
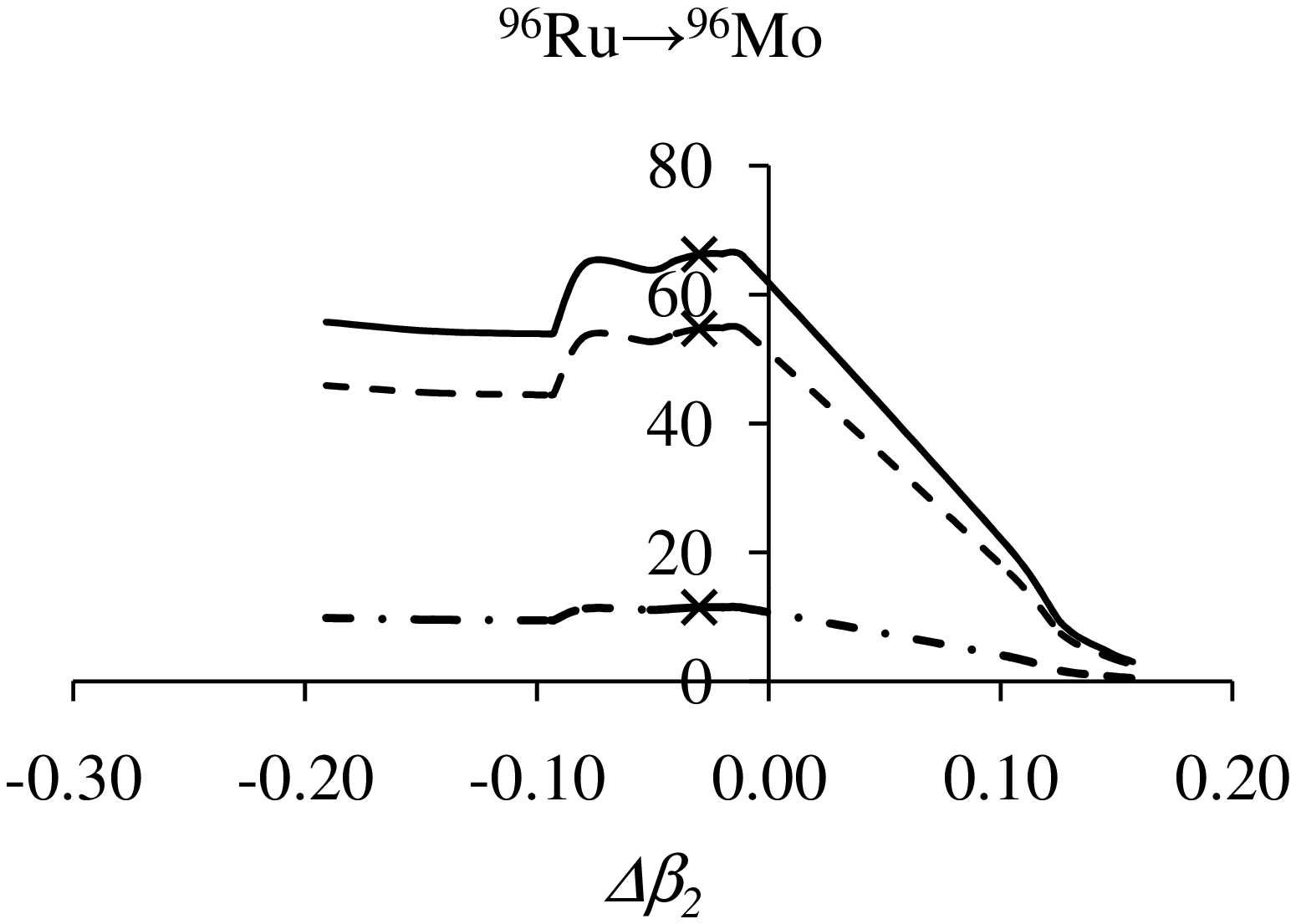} \\ 
\includegraphics [scale=0.40]{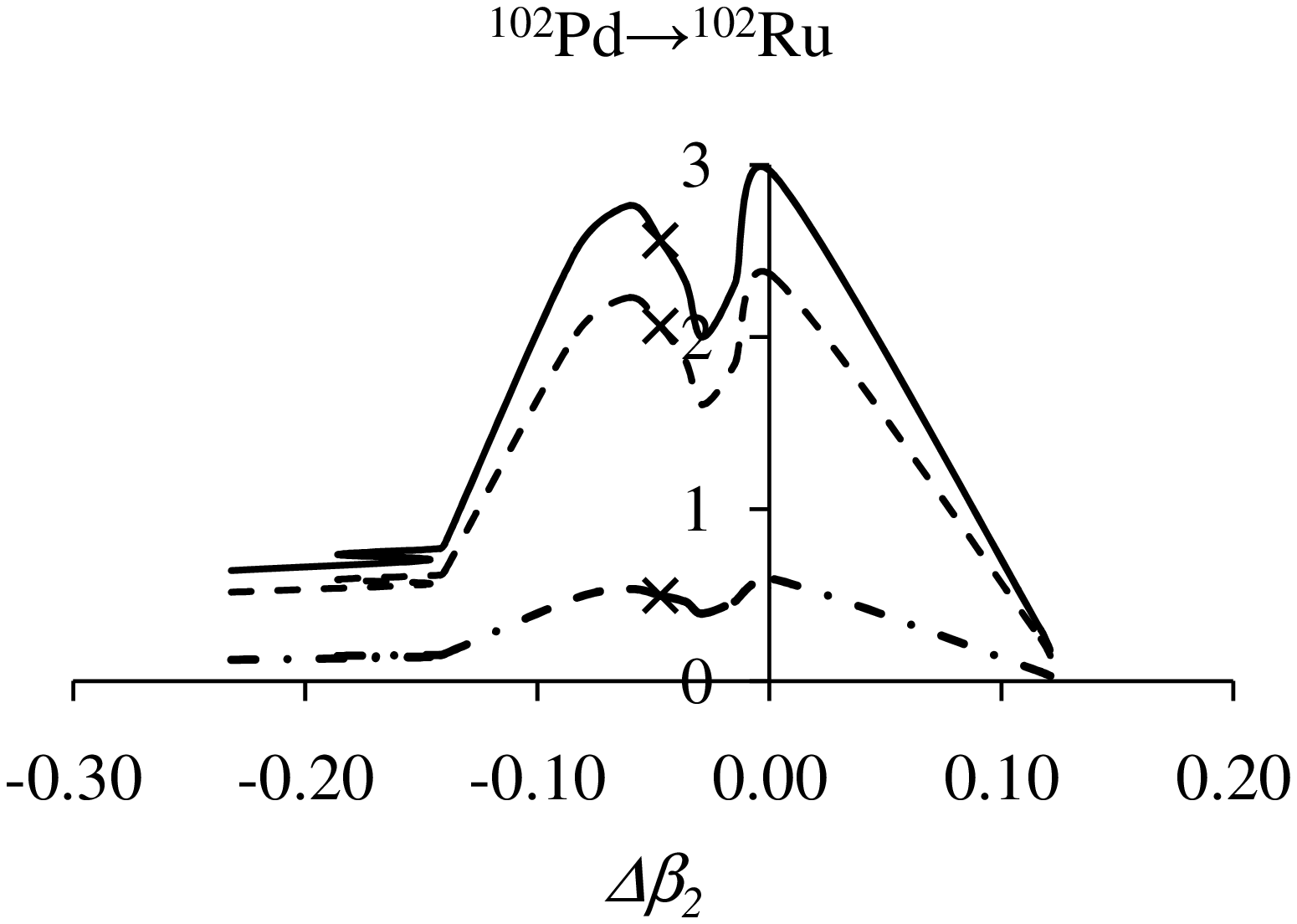} & \includegraphics [scale=0.40]{%
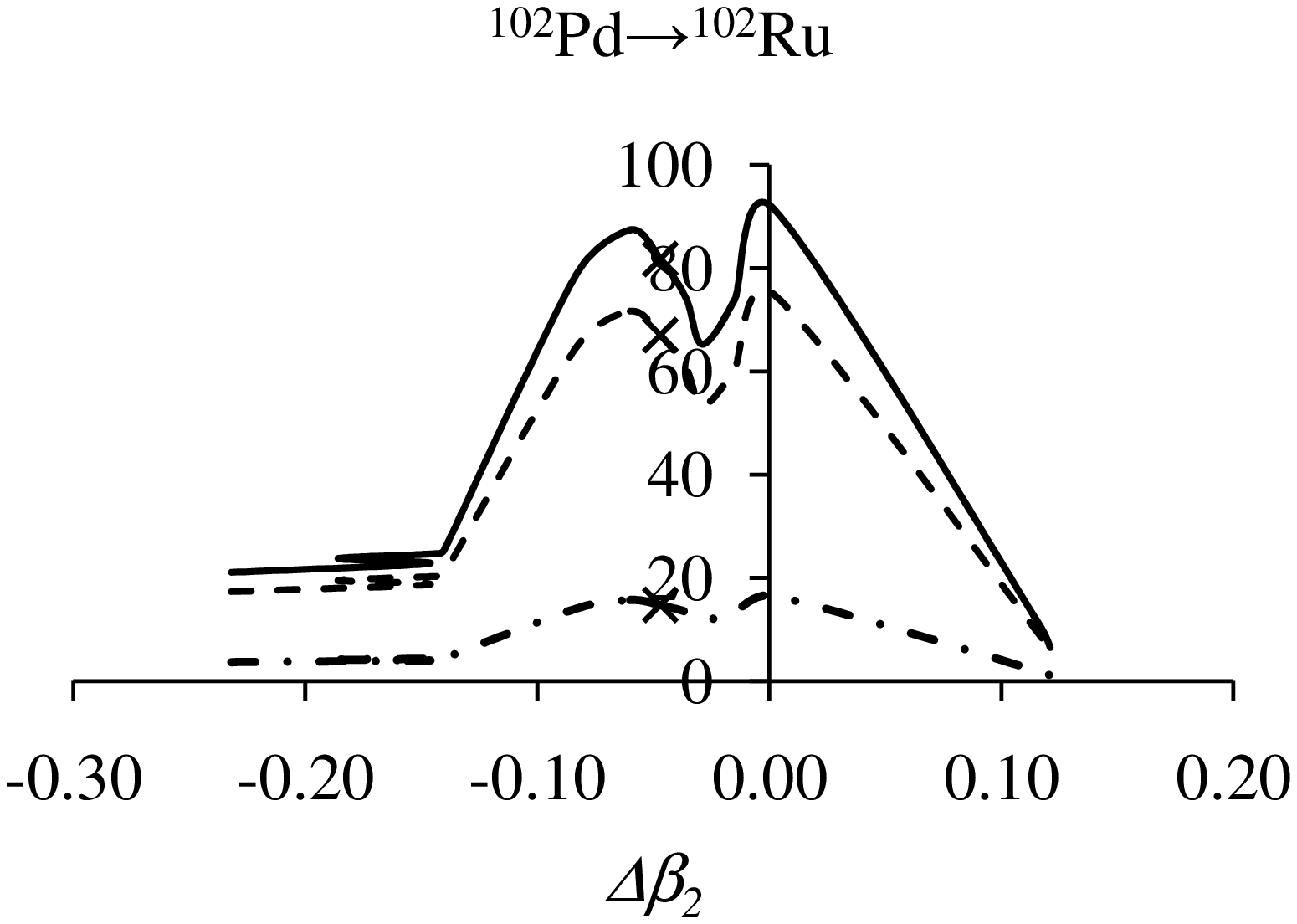} \\ 
\includegraphics [scale=0.40]{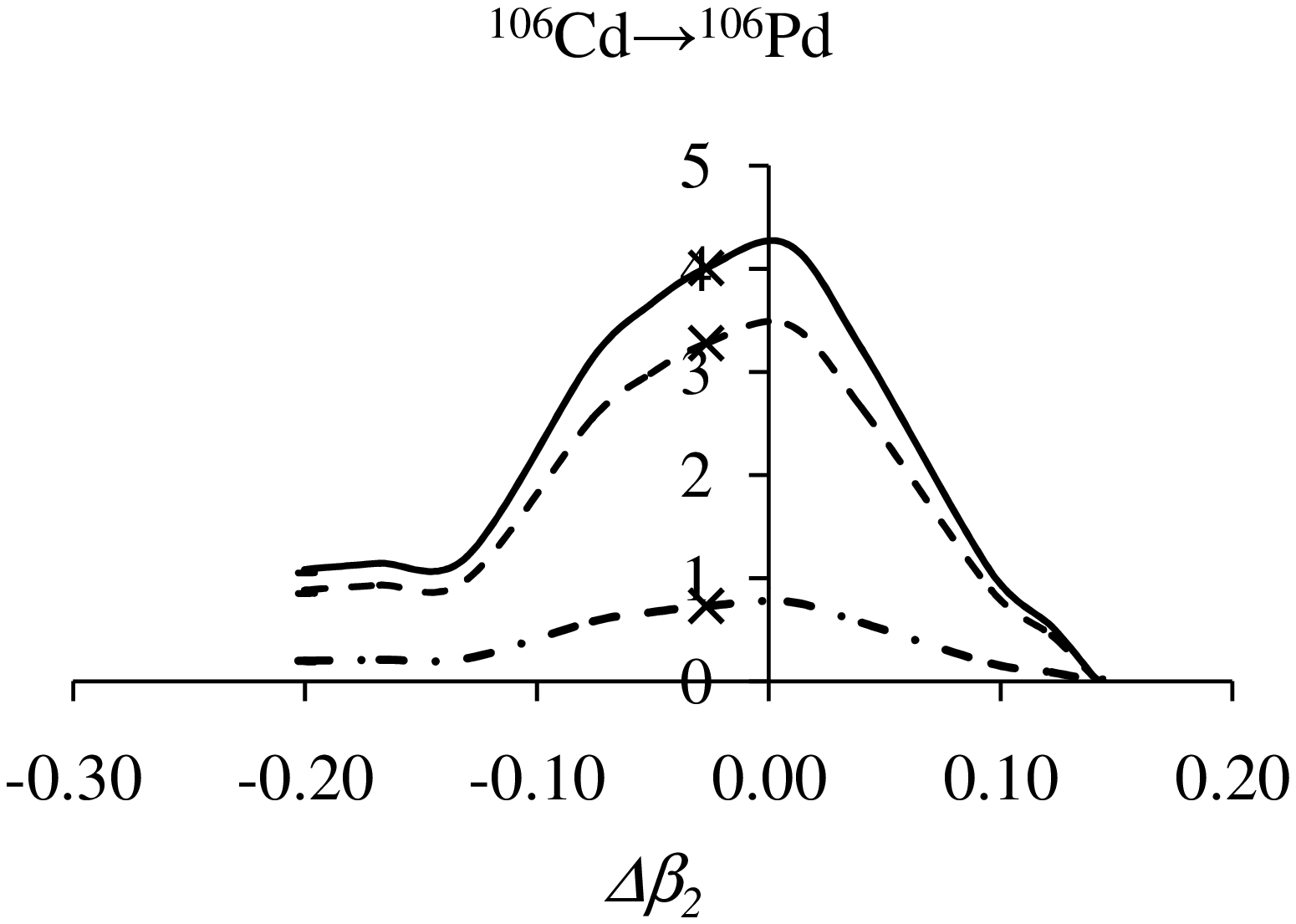} & \includegraphics [scale=0.40]{%
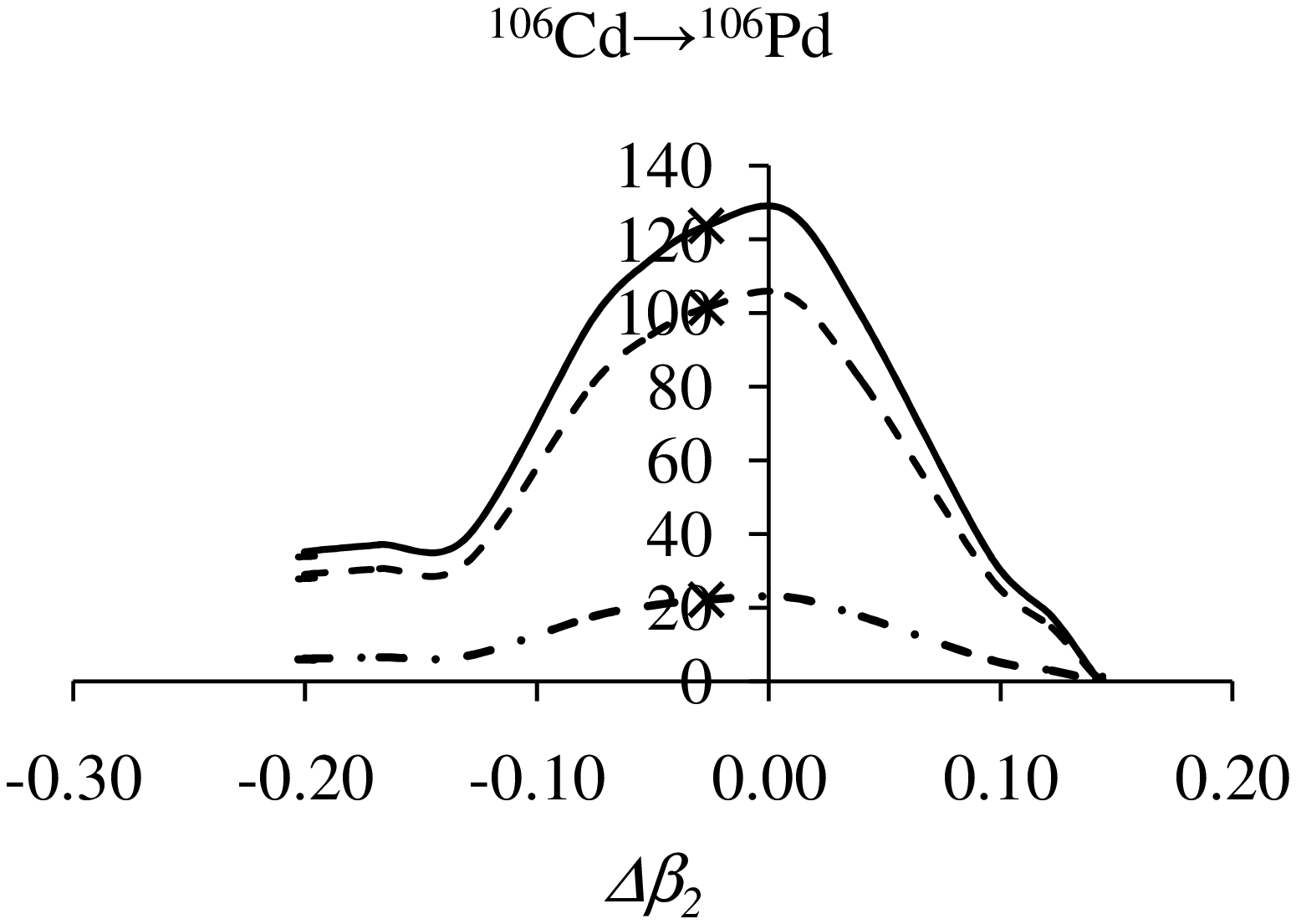} \\ 
\end{tabular}
\caption{NTMEs 
of $\left(\beta^{+}\beta^{+}\right)_{0\nu}$
and $\left(\varepsilon\beta^{+}\right)_{0\nu}$ modes
for $^{96}$Ru, $^{102}$Pd, $^{106}$Cd isotopes due to the exchange 
of light (left hand side) 
and heavy (right hand side) neutrinos as a function of the difference 
in the deformation parameters $%
\Delta\beta_2$. ``$\times$" denotes the NTME for calculated $\Delta\beta_2$
at $\zeta_{qq}=1$.}
\label{fig2}
\end{figure*}

\begin{figure*}[htb]
\begin{tabular}{cc}
\includegraphics [scale=0.40]{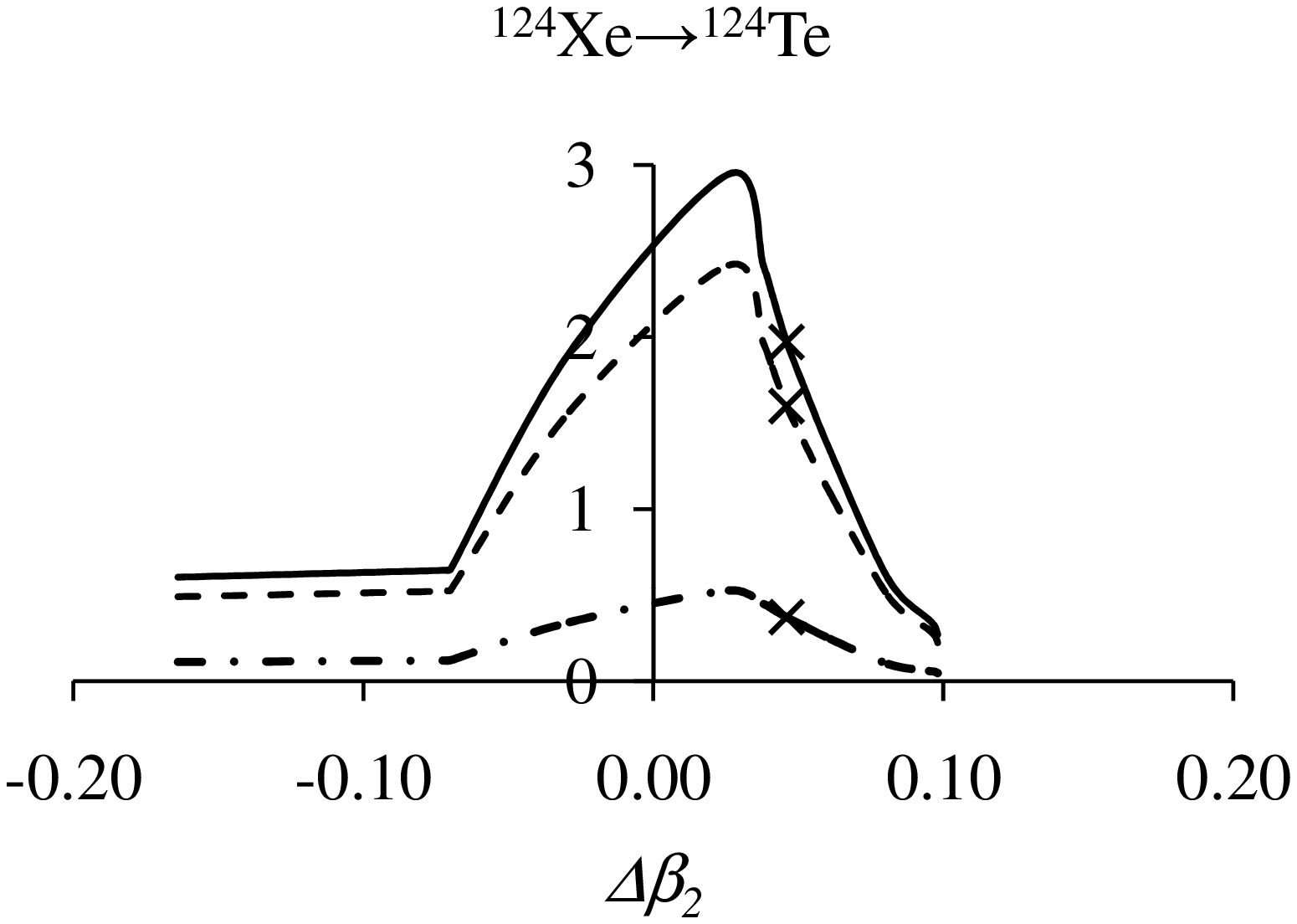} & \includegraphics [scale=0.40]{%
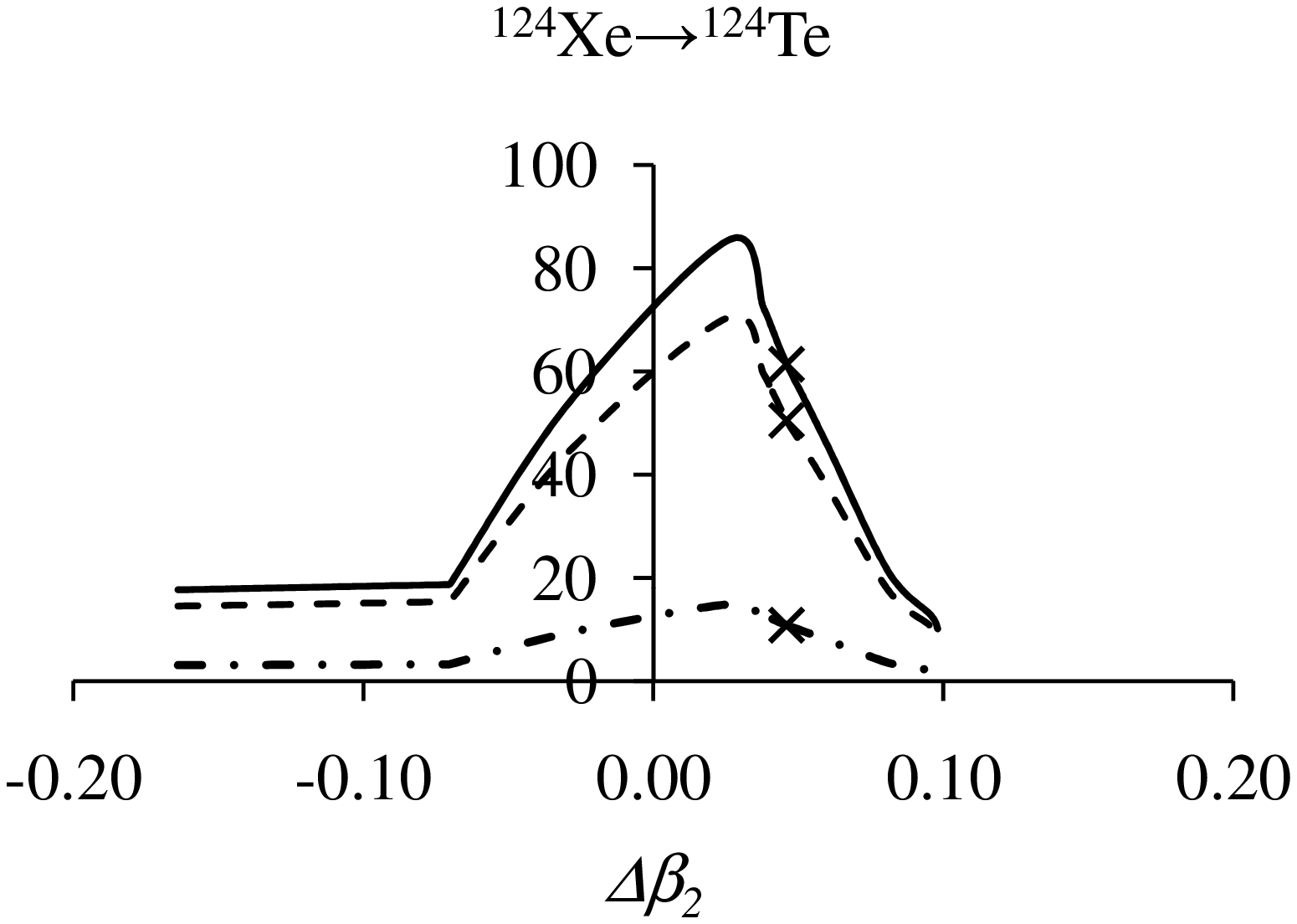} \\
\includegraphics [scale=0.40]{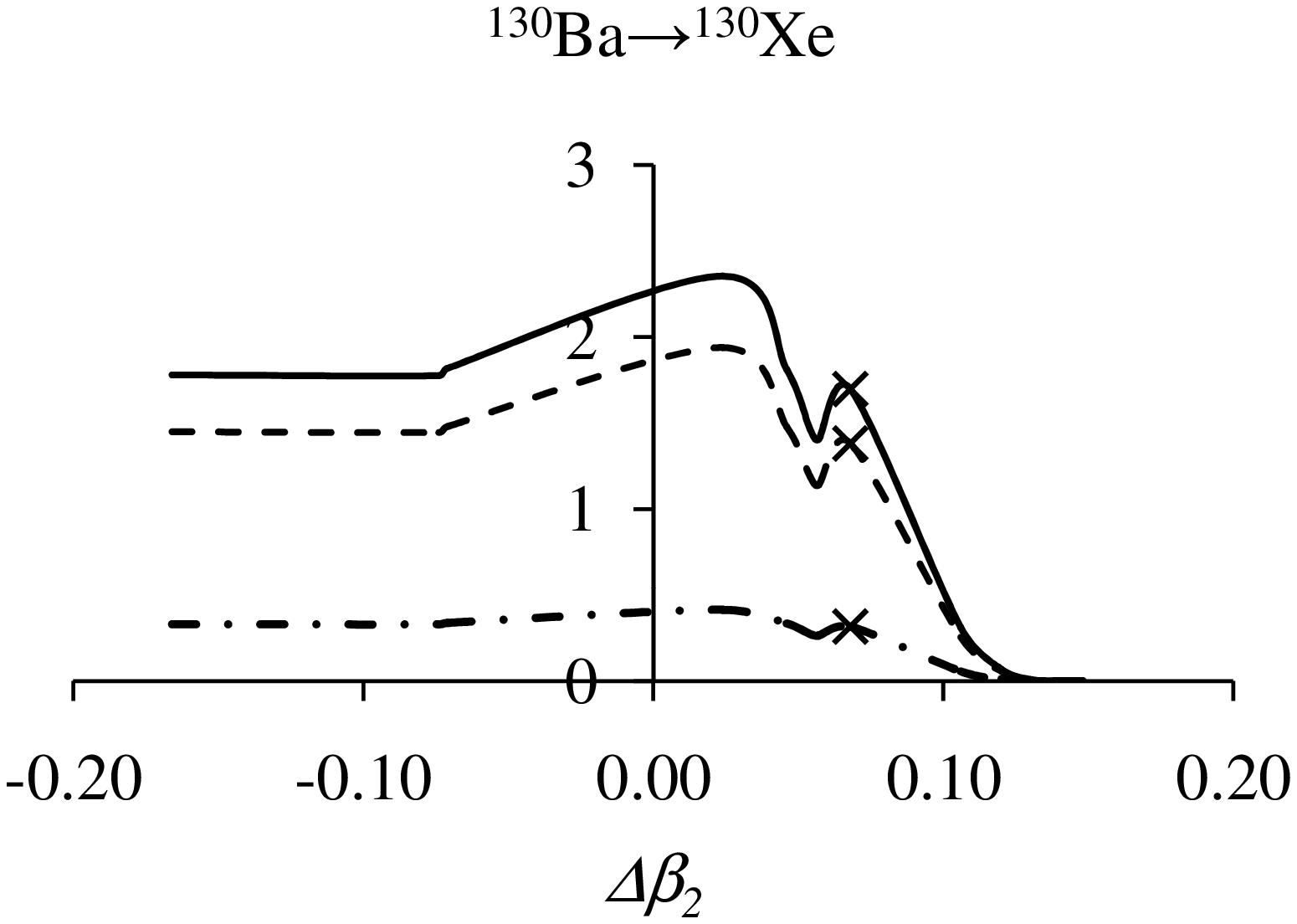} & \includegraphics [scale=0.40]{%
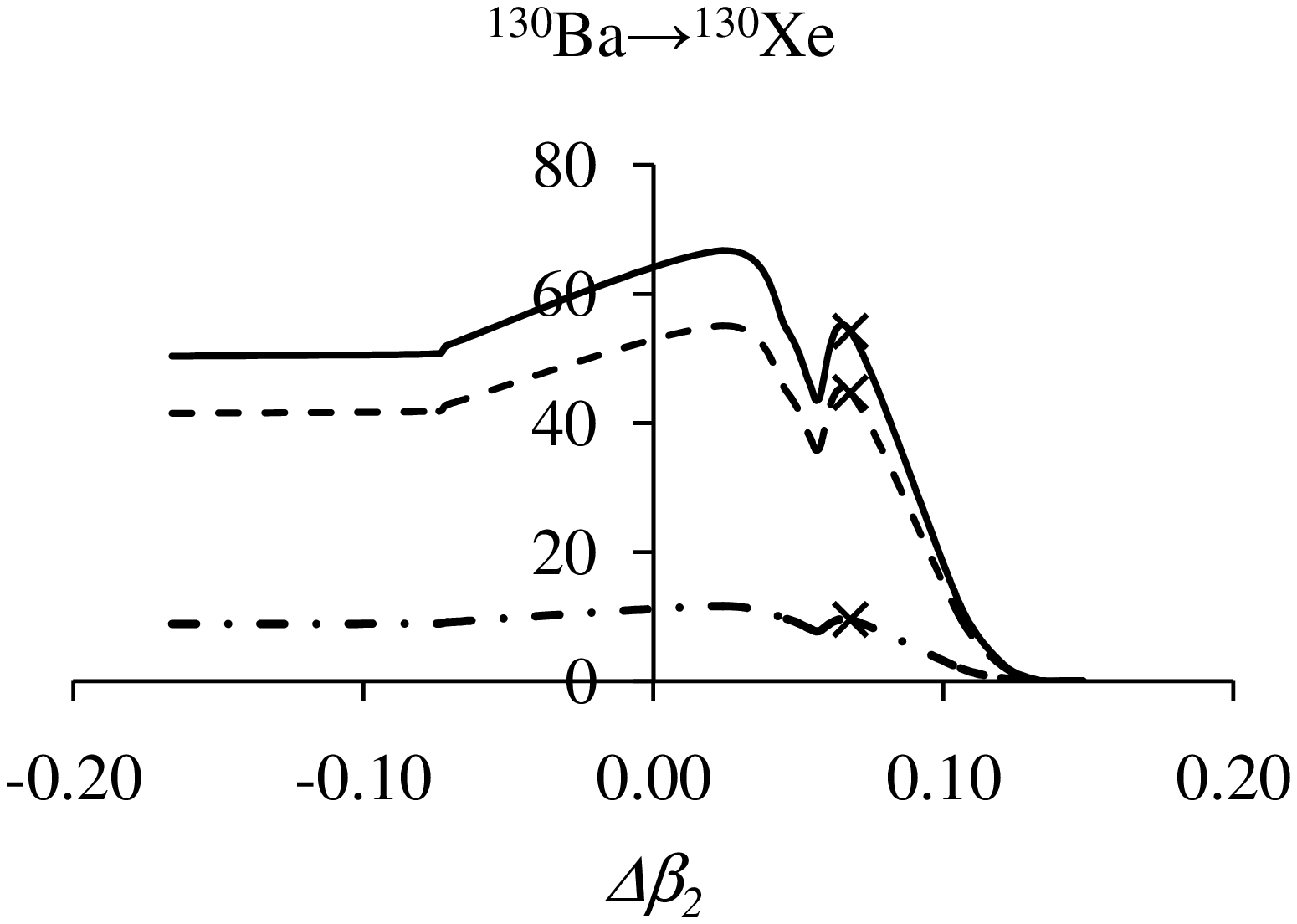} \\
\includegraphics [scale=0.40]{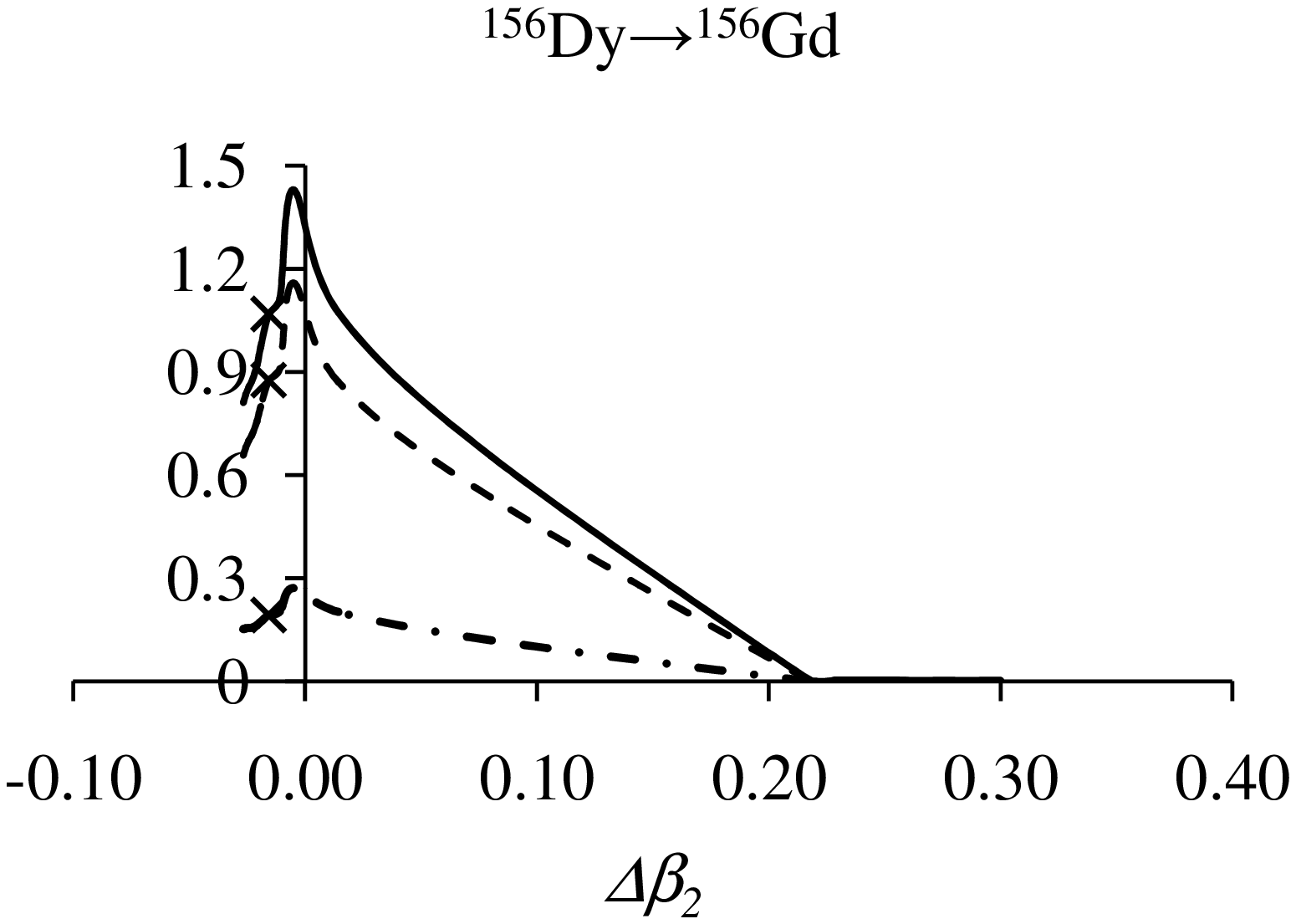} & \includegraphics [scale=0.40]{%
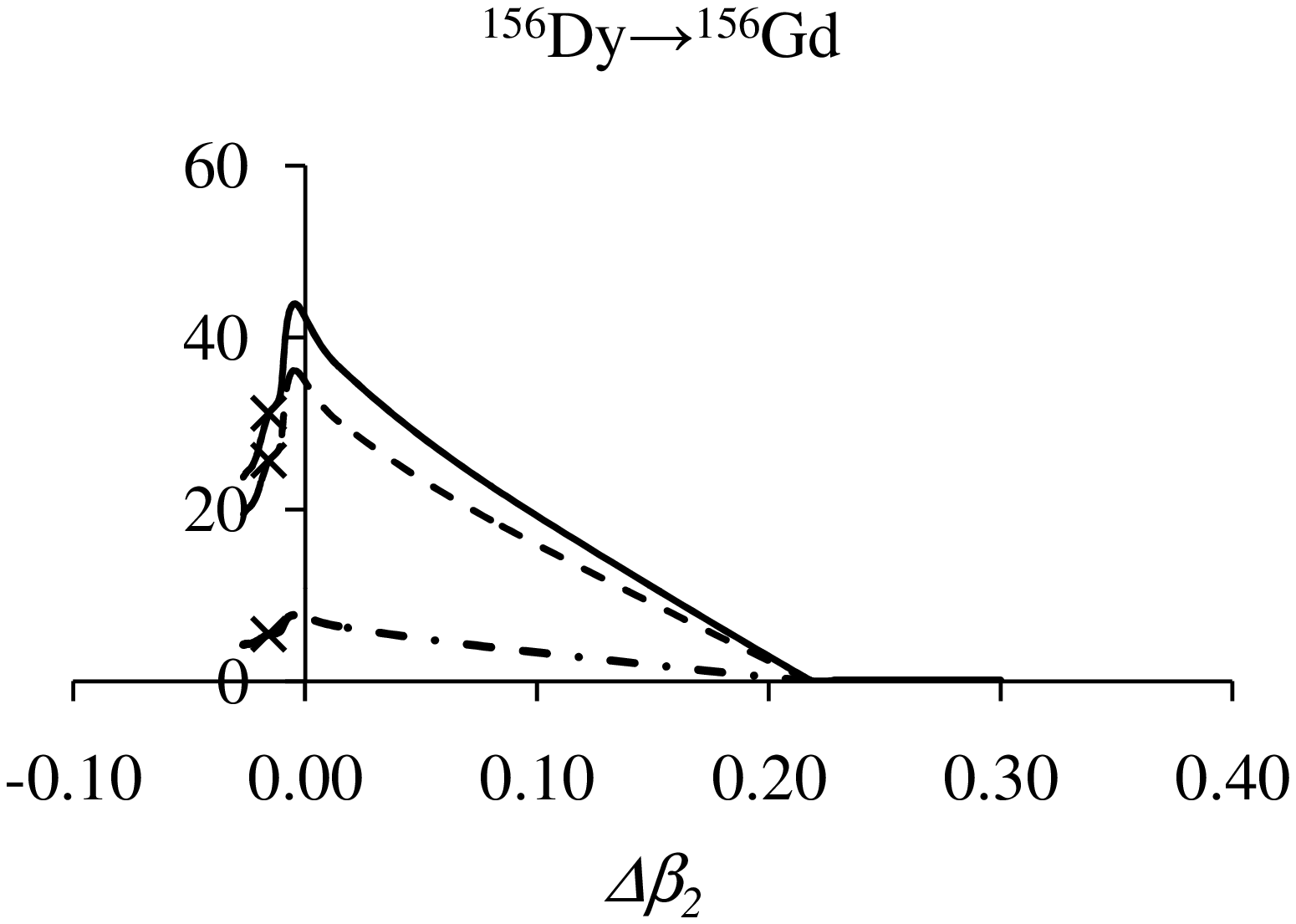}
\end{tabular}
\caption{NTMEs of $\left(\beta^{+}\beta^{+}\right)_{0\nu}$
and $\left(\varepsilon\beta^{+}\right)_{0\nu}$ modes
for $^{124}$Xe,$^{130}$Ba and $^{156}$Dy isotopes. Further details are
given in Fig.~\ref{fig2}.}
\label{fig3}
\end{figure*}

In the left and right panels of Fig.~\ref{fig2} and \ref{fig3}, we 
present the variation of NTMEs $%
\left| M_{0\nu }\right| $ and $\left| M_{0N}\right| $ due to the light and
heavy neutrino exchange, respectively, with respect to $\Delta \beta
_{2}=\beta _{2}(parent)-$ $\beta _{2}(daughter)$ for the above mentioned $%
e^{+}\beta \beta $ emitters. The theoretically calculated deformation parameters 
$\beta_2$ for parent and daughter nuclei have been given in Refs. \cite{rain06,sing07} and we
present them in Table~\ref{tab6} for convenience. It 
can be
noticed that the variation in $\left|
M_{0\nu }\right| $ with changing $\Delta \beta _{2}$ is similar as that of $%
\left| M_{0N}\right| $. Moreover, it 
can be observed in
Fig.~\ref{fig2} and \ref{fig3} that the NTMEs
remain constant even when one of the nuclei is spherical or slightly
deformed. With further increase in deformation, the NTMEs in general become the maximum
for $\Delta \beta _{2}=0$ and then decrease with increase in the difference
between the deformation parameters. To summarize, the independent
deformations of initial and final nuclei are important parameters to
describe the NTMEs $M_{0\nu }$ and $M_{0N}$ of $\left( \beta ^{+}\beta
^{+}\right) _{0\nu }$ and $\left( \varepsilon \beta ^{+}\right) _{0\nu }$ modes.

\begin{table}[htb]
\caption{Ratios $D_{\alpha }$ for $^{96}$Ru, $^{102}$Pd, $^{106}$Cd, $^{124}$%
Xe, $^{130}$Ba and $^{156}$Dy isotopes.}
\label{tab5}
\begin{tabular}{lllllll}
\hline\hline
Ratios & $^{96}$Ru\thinspace \thinspace \thinspace \thinspace & $^{102}$%
Pd\thinspace \thinspace \thinspace \thinspace & $^{106}$Cd\thinspace
\thinspace \thinspace \thinspace & $^{124}$Xe\thinspace \thinspace
\thinspace \thinspace & $^{130}$Ba\thinspace \thinspace \thinspace \thinspace
& $^{156}$Dy \\ \hline
&  &  &  &  &  &  \\ 
$D_{F}$ & 2.92 & 2.52 & 1.91 & 3.83 & 4.68 & 10.42 \\ 
&  &  &  &  &  &  \\ 
$D_{GT}$ & 2.48 & 2.73 & 1.96 & 3.88 & 4.72 & 10.68 \\ 
&  &  &  &  &  &  \\ 
$D_{Fh}$ & 2.61 & 2.34 & 1.72 & 3.42 & 4.11 & 10.20 \\ 
&  &  &  &  &  &  \\ 
$D_{GTh}$ & 2.49 & 2.36 & 1.72 & 3.45 & 4.13 & 10.20 \\ 
&  &  &  &  &  &  \\ 
$D_{2\nu }$ & 3.13 & 3.40 & 2.06 & 3.63 & 4.66 & 13.64 \\ \hline\hline
\end{tabular}
\end{table}
\begin{table}[htb]
\caption{Calculated \cite{rain06,sing07} and experimental \cite{rama01}
deformation parameters $\beta_2$ of parent and daughter
nuclei participating in $\left(\beta^{+}\beta^{+}\right)_{0\nu}$
and $\left(\varepsilon\beta^{+}\right)_{0\nu}$ modes.}
\label{tab6}
\begin{tabular}{ccr}
\hline\hline
Nuclei & \multicolumn{2}{c}{$~~~~~\beta _{2}$} \\
&~~~~~~~~~~~~~~Theory~~~~~~~~~~~~~~& Experiment \\ \hline
$^{96}$Ru &~~~~~~~~~~~~~~0.161~~~~~~~~~~~~~~& 0.1579$\pm 0.0031$ \\
$^{96}$Mo &~~~~~~~~~~~~~~0.191~~~~~~~~~~~~~~& 0.1720$\pm 0.0016$ \\
$^{102}$Pd &~~~~~~~~~~~~~~0.185~~~~~~~~~~~~~~& 0.196$\pm 0.006$ \\
$^{102}$Ru &~~~~~~~~~~~~~~0.232~~~~~~~~~~~~~~& 0.2404$\pm 0.0019$ \\
$^{106}$Cd &~~~~~~~~~~~~~~0.176~~~~~~~~~~~~~~& 0.1732$\pm 0.0042$ \\
$^{106}$Pd &~~~~~~~~~~~~~~0.203~~~~~~~~~~~~~~& 0.229$\pm 0.006$ \\
$^{124}$Xe &~~~~~~~~~~~~~~0.210~~~~~~~~~~~~~~& 0.212$\pm 0.007$ \\
$^{124}$Te &~~~~~~~~~~~~~~0.164~~~~~~~~~~~~~~& 0.1695$\pm 0.0009$ \\
$^{130}$Ba &~~~~~~~~~~~~~~0.234~~~~~~~~~~~~~~& 0.2183$\pm 0.0015$ \\
$^{130}$Xe &~~~~~~~~~~~~~~0.166~~~~~~~~~~~~~~& 0.169$\pm 0.007$ \\
$^{156}$Dy &~~~~~~~~~~~~~~0.300~~~~~~~~~~~~~~& 0.2929$\pm 0.0016$ \\
$^{156}$Gd &~~~~~~~~~~~~~~0.316~~~~~~~~~~~~~~& 0.3378$\pm 0.0018$ \\ \hline\hline
\end{tabular}
\end{table}
\section{CONCLUSIONS}

We 
have calculated
the NTMEs $M_{F}$, $M_{GT}$, $M_{Fh}$ and $M_{GTh}$ required to
study the $\left( \beta ^{+}\beta ^{+}\right) _{0\nu }$ mode of $^{96}$Ru, $%
^{106}$Cd, $^{124}$Xe and $^{130}$Ba as well as the $\left( \varepsilon
\beta ^{+}\right) _{0\nu }$ mode of $^{96}$Ru, $^{102}$Pd, $^{106}$Cd, $%
^{124}$Xe, $^{130}$Ba and $^{156}$Dy nuclei for the $0^{+}\rightarrow 0^{+}$
transition in the Majorana neutrino mass mechanism using the set of HFB wave 
functions, the
reliability of which was tested by obtaining an overall agreement between
theoretically calculated results for the yrast spectra, reduced $B(E2$:$%
0^{+}\rightarrow 2^{+})$ transition probabilities, static quadrupole moments 
$Q(2^{+})$ and $g$-factors $g(2^{+})$ and NTMEs $M_{2\nu }$ as well as
half-lives $T_{1/2}^{2\nu }$ of $\left( e^{+}\beta \beta \right) _{2\nu }$
decay and the available experimental data \cite{rain06,sing07,rath09}. The
existing experimental data on $\left( \beta ^{+}\beta ^{+}\right) _{0\nu }$
and $\left( \varepsilon \beta ^{+}\right) _{0\nu }$ modes fail to provide
stringent limits on the extracted effective mass of light neutrino $%
\left\langle m_{\nu }\right\rangle $ and heavy neutrino $\left\langle
M_{N}\right\rangle $. Hence, we calculate half-lives $T_{1/2}^{0\nu }$ of
these modes for the light neutrino and extract limits on 
$\left\langle M_{N}\right\rangle $. In the
mass mechanism, the half-lives $T_{1/2}^{0\nu }\left( \beta ^{+}\beta
^{+}\right) $ and $T_{1/2}^{0\nu }\left( \varepsilon \beta ^{+}\right) $ are
related through the exactly calculable phase space factors $G_{01}\left(
\beta ^{+}\beta ^{+}\right) $ and $G_{01}\left( \varepsilon \beta
^{+}\right) $. In addition, it is observed that the ratio of NTMEs $\left|
M_{0N}\right| /\left| M_{0\nu }\right| $ $\approx 30$ is a 
constant for different nuclei
so that half-lives due to the exchange of light and heavy neutrinos are
also in constant ratio. Further, the role of deformation on NTMEs $%
M_{F}, $ $M_{GT}$, $M_{Fh}$ and $M_{GTh}$ for $\left( \beta ^{+}\beta
^{+}\right) _{0\nu }$ and $\left( \varepsilon \beta ^{+}\right) _{0\nu }$
modes is investigated by changing the strength $\zeta _{qq}$ of the \textit{%
QQ} interaction. It is noticed that there is an inverse correlation between
the magnitudes of NTMEs and quadrupole moments $Q(2^{+})$ as well as deformation
parameters $\beta _{2}$. The NTMEs are suppressed by factors of 1.7--10.7
in the considered mass range $A=96-156$ implying that the nuclear structure
effects are also important for $\left( \beta ^{+}\beta ^{+}\right) _{0\nu }$
and $\left( \varepsilon \beta ^{+}\right) _{0\nu }$ modes. The deformation
of individual nucleus is an important parameter for calculating NTMEs $%
M_{0\nu }$ and $M_{0N}$ of $\left( \beta ^{+}\beta ^{+}\right) _{0\nu }$ and 
$\left( \varepsilon \beta ^{+}\right) _{0\nu }$ modes.

This work has been partially supported by DST, India vide grant No.
SR/S2/HEP-13/2006, by Conacyt-M\'{e}xico and DGAPA-UNAM.

\end{document}